\documentstyle[thmsa,a4,sw20lart]{article}

\input tcilatex
\QQQ{Language}{
British English
}

\begin{document}

\section{Introduction}

In a recent paper, we have introduced a new treatment of systems of
compounded angular momentum which leads to very generalized formulas for the
states and operators for such systems [1]. In the paper, we worked these
quantities out explicitly for the cases of spin $0$ and spin $1$ resulting
from the addition of two spins of $1/2$ each. We first obtained the
generalized probability amplitudes describing results of measurements on
such systems, then used these to derive the matrix treatment of the systems.
The forms that we obtained for the vector states and operators proved to be
entirely different from the standard forms. However, as might be expected,
the results of calculations of measurable quantities are the same.
Nevertheless, the question how the vectors and operators belonging to the
standard treatment of spin addition are related to the new forms requires an
answer. Indeed, since we consider the treatment of spin addition by means of
probability amplitudes as being the foundation of any matrix treatment, it
is necessary to derive the standard quantities by the new approach.

In this paper, we demonstrate that the standard matrix treatment of
compounded spin is indeed derivable by our method. The use of this approach
not only yields the standard treatment, but also produces results more
generalized than any in the literature. However, these results reduce to the
standard forms in an appropriate limit. The systems on which the theory
developed is tested are the triplet and singlet states resulting from the
addition of the spins of two spin-$1/2$ systems. It is clear from the
application of the theory to these cases how the extension to arbitrary
systems of compounded spin is achieved.

The organization of the paper is as follows. After the introduction in
Section $1$, we give in Section $2$ a brief description of those features of
the Land\'e approach to quantum mechanics that we shall use to develop our
treatment. In Section $3$, we turn our attention to a review of the work we
have so far done on systems of compounded angular momentum. We remind
ourselves of the expressions for the probability amplitudes for the addition
of general angular momentum in Section $3.1$, and of the probability
amplitudes for spin addition in Section $3.2$.

In Section $4$, we look at the way the transformation from wave or
probability- amplitude mechanics to matrix mechanics is achieved. We sketch
in Section $4.1$ the derivation of matrix mechanics from
probability-amplitudes mechanics for simple systems. In Section $4.2$, we
derive the standard form of matrix mechanics for systems of compounded spin
- however, the new results are more generalized than the standard ones.

The results of Section $4$ are employed on actual systems in Section $5$.
The test systems are the singlet and triplet states arising from the
addition of the spins of two spin -$1/2$ systems. The matrix operator is
common to both cases, and is calculated in Section $5.1$. The vectors states
are obtained in Section $5.2.$

The results obtained in Section $5$ are more generalized than the standard
forms found in the literature. In Section $6$ we demonstrate how to reduce
these results to the standard forms.

We end the paper with a Discussion and Conclusion in Section $7$.

\section{Basic Theory}

\subsection{The Land\'e Approach to Quantum Mechanics}

The basic theory underlying our work derives from the interpretation of
quantum mechanics due to Land\'e [2-5]. Among many features of the Land\'e
approach is the assumption that wave functions and eigenfunctions in quantum
mechanics are probability amplitudes. Any probability amplitude connects two
states - one state pertaining to the situation that obtains before a
measurement is made, and the other to the state that results from the
measurement. Thus, an energy eigenfunction $\phi _E({\bf r})$ resulting from
solution of the time-independent Schr\"odinger equation is a probability
amplitude connecting two states: the state defined by the eigenvalue $E$ is
the initial state, while the final state is characterized by the position
eigenvalue ${\bf r}$. Thus $\left| \phi _E({\bf r})\right| ^2d{\bf r}$ is
the probability that if the system is initially in the state corresponding
to $E$, a measurement of the position gives ${\bf r}$ in the volume element $%
d{\bf r}$.

The Land\'e interpretation of quantum mechanics is based on the principle
that nature is ultimately indeterministic and should be described by a
theory that is fundamentally probabilistic. Consequently, the description of
measurements can only be given in probabilistic terms through probability
amplitudes. For a particular system, the different sets of probability
amplitudes connecting different measurable quantities are inter-related in
the following way.

Let a quantum system have the observables $A$, $B$ and $C$ which have the
respective eigenvalue spectra $A_1$, $A_2,$...,$A_N,$ $B_1$, $B_2,$...,$B_N$
and $C_1$, $C_2,..,C_N.$ If the system is initially in the state
corresponding to the eigenvalue $A_i$, a measurement of $B$ yields any of
the eigenvalues $B_j$ with probabilities determined by the probability
amplitudes $\eta (A_i;B_j).$ A measurement of $C$ results in one of the
eigenvalues $C_j$ with probabilities determined by the probability
amplitudes $\psi (A_i;C_j).$ If the system is initially in the state
corresponding to the eigenvalue $B_i$, a measurement of $C$ gives any of the
eigenvalues $C_j$ with probabilities determined by the probability
amplitudes $\xi (B_i;C_j).$ The probability amplitudes display a two-way
symmetry contained in the Hermiticity condition

\begin{equation}
\psi (C_j;A_i)=\psi ^{*}(A_i;C_j).  \label{on1}
\end{equation}

These probability amplitudes are orthogonal:

\begin{equation}
\sum_{j=1}^N\psi ^{*}(A_i;C_j)\psi (A_k;C_j)=\delta _{ik}.  \label{fo4p}
\end{equation}

The law that connects the three sets of probability amplitudes is 
\begin{equation}
\psi (A_i;C_n)=\sum_{j=1}^N\eta (A_i;B_j)\xi (B_j;C_n).  \label{tw2}
\end{equation}

Though the features of the Land\'e approach highlighted above refer to
probability amplitudes that correspond to a discrete final eigenvalue
spectrum, there is no essential difference if this spectrum is instead
continuous. In fact, if the observable $C$ is the position ${\bf r}$, and
if, as is customary, we ignore the initial state $A_i$ in the labelling, Eq.
(\ref{tw2}) becomes 
\begin{equation}
\psi ({\bf r})=\sum_{j=1}^N\eta _j\xi _j({\bf r}),  \label{tw2a}
\end{equation}
where we have set $\eta _j=\eta (B_j)$ and $\xi _j=\xi (B_j;C_n).$ We
recognize this equation as the law of interference of probabilities. In the
Land\'e formalism, this important relation is derived, not assumed.

The relation Eq. (\ref{tw2a}) is, of course, the basis for the
transformation of representation from wave to matrix mechanics for the case
where the eigenfunctions $\xi _j({\bf r})$ are known from solution of some
eigenvalue equation. By the same token, its parent relation Eq. (\ref{tw2})
is the basis for the transformation of representation from
probability-amplitude mechanics to matrix mechanics in all cases,
irrespective of whether or not a differential eigenvalue equation exists for
the probability amplitudes $\xi (B_j;C_n).$ Indeed, this is the relation on
which we have based the derivation of the matrix theory of spin from
probability amplitudes [1,7-11].

\section{Review of Previous Results on Angular Momentum Addition}

\subsection{General Theory}

In a previous paper [1], we derived a matrix treatment of spin addition
which resulted in new forms for the vectors and the operators, apart from
throwing light on the theory of angular momentum addition. In this section,
we review the new treatment of spin addition in order to present those
results which will be needed in the development of the present work.

We consider first the case of general angular momentum addition. Let a
system have the total angular momentum ${\bf J}$ resulting from adding the
angular momenta ${\bf J}_1$ and ${\bf J}_2$ of subsystems $1$ and $2$. Thus, 
\begin{equation}
{\bf J}={\bf J}_1+{\bf J}_2.  \label{el11}
\end{equation}
The quantum numbers of the angular momenta of the subsystems are $j_1$ and $%
j_2$, while that of the angular momentum of the total system is $j$. The $z$
components of these respective angular momenta are characterized by the
quantum numbers $m_1$, $m_2$ and $M$. For the time being, we shall assume
that ${\bf J,}$ ${\bf J}_1$ and ${\bf J}_2$ are orbital angular momenta. The
subsystems $1$ and $2$ are characterized by the angular variables $(\theta
_1,\varphi _1)$ and $(\theta _2,\varphi _2)$ respectively. The standard
expression for the wave function of the coupled system is

\begin{equation}
\Psi _{j_1j_2jM}(\theta _1,\varphi _1,\theta _2,\varphi
_2)=\sum_{m_1}C(j_1j_2j;m_1m_2M)\phi _{j_1m_1}^{(1)}(\theta _1,\varphi
_1)\phi _{j_2m_2}^{(2)}(\theta _2,\varphi _2),  \label{tw12}
\end{equation}
where we have used the notation in Rose [6] for the Clebsch-Gordan
coefficients $C(j_1j_2j;m_1m_2M)$. If we are dealing with orbital angular
momentum, the $\phi _{j_1m_1}^{(1)}(\theta _1,\varphi _1)$ and the $\phi
_{j_2m_2}^{(2)}(\theta _2,\varphi _2)$ are spherical harmonics.

In the Land\'e interpretation, the function $\Psi _{j_1j_2jM}(\theta
_1,\varphi _1,\theta _2,\varphi _2)$ is a probability amplitude. Its
expression in terms of an expansion must be of the general structure of Eq. (%
\ref{tw2}). Therefore, we rewrite Eq. (\ref{tw12}) in the following way: 
\begin{eqnarray}
&&\Psi (j_1,j_2,j,M;\theta _1,\varphi _1,\theta _2,\varphi
_2)=\sum_{m_1}\chi (j_1,j_2,j,M;j_1,m_1,j_2,m_2)  \nonumber \\
&&\times \Phi (j_1,m_1,j_2,m_2;\theta _1,\varphi _1,\theta _2,\varphi _2),
\label{th13}
\end{eqnarray}
where 
\begin{equation}
\chi (j_1,j_2,j,M;j_1,m_1,j_2,m_2)=C(j_1j_2j;m_1m_2M)  \label{th13a}
\end{equation}
and 
\begin{equation}
\Phi (j_1,m_1,j_2,m_2;\theta _1,\varphi _1,\theta _2,\varphi _2)=\phi
_{j_1m_1}^{(1)}(\theta _1,\varphi _1)\phi _{j_2m_2}^{(2)}(\theta _2,\varphi
_2).  \label{th13b}
\end{equation}
Then the various quantities have the following interpretations:

The function $\Psi (j_1,j_2,j,M;\theta _1,\varphi _1,\theta _2,\varphi _2)$
is a probability amplitude characterized by an initial state corresponding
to the quantum numbers $(j_1,j_2,j,M)$, and a final state corresponding to
the eigenvalues $(\theta _1,\varphi _1,\theta _2,\varphi _2).$ In the
initial state, $j$ is the quantum number for the total angular momentum, $%
M\hbar $ is the projection of the total angular momentum along the $z$
direction, $j_1$ is the quantum number of subsystem $1$ and $j_2$ is the
quantum number of subsystem $2$. In the state which results from the
measurement, $(\theta _1,\varphi _1)$ is the angular position of subsystem
1, while $(\theta _2,\varphi _2)$ is the angular position of subsystem $2$.
Thus, this probability amplitude gives the probability for obtaining
specified angular positions of systems $1$ and $2$ upon measurement if the
initial state of the compound system is defined by the quantum numbers $%
(j_1,j_2,j,M)$.

The Clebsch-Gordan coefficient $\chi (j_1,j_2,j,M;j_1,m_1,j_2,m_2)$ is a
probability amplitude characterized by an initial state corresponding to $%
(j_1,j_2,j,M)$ and a final state defined by $(j_1,m_1,j_2,m_2)$. In the
state resulting from the measurement, the angular momentum quantum number of
subsystem $1$ is $j_1$, while its component in the $z$ direction is $%
m_1\hbar $, and the angular momentum quantum number of subsystem $2$ is $j_2$%
, while its $z$ projection is $m_2\hbar $. This probability amplitude thus
gives the probability of obtaining specified projections of the angular
momenta of the subsystems along the $z$ axis starting from a state of the
compound system defined by the quantum numbers $(j_1,j_2,j,M).$

The function $\Phi (j_1,m_1,j_2,m_2;\theta _1,\varphi _1,\theta _2,\varphi
_2)$ is a probability amplitude with an initial state defined by $%
(j_1,m_1,j_2,m_2)$ and a final state defined by the eigenvalues $(\theta
_1,\varphi _1,\theta _2,\varphi _2).$ This probability amplitude thus gives
the probability of obtaining specified angular positions of the subsystems
starting from a state characterized by specified projections of these
subsystems along the $z$ direction.

In labelling the various probability amplitudes, we may reduce on the
clutter by suppressing those quantum numbers which do not change at all
during the measurement. Thus, we omit $j_1$ and $j_2$. However, we retain
the subscript $j$ because for given $j_1$ and $j_2$, several values of $j$
are possible within the limits 
\begin{equation}
j_1+j_2\leq j\leq \left| j_1-j_2\right| .  \label{th13f}
\end{equation}

With these changes, Eq. (\ref{th13}) becomes

\begin{equation}
\Psi (j,M;\theta _1,\varphi _1,\theta _2,\varphi _2)=\sum_{m_1}\chi
(j,M;m_1,m_2)\Phi (m_1,m_2;\theta _1,\varphi _1,\theta _2,\varphi _2).
\label{th13c}
\end{equation}

We have elsewhere interpreted the probability amplitude $\Psi (j,M;\theta
_1,\varphi _1,\theta _2,\varphi _2)$ as a special form of the probability
amplitude $\Psi (j(\theta ,\varphi ),M;\theta _1,\varphi _1,\theta
_2,\varphi _2)$[1]$.$ The former quantity is specialized because it pertains
to a situation where projections of the total angular momentum are measured
with respect to the $z$ direction (for which $\theta =\varphi =0$), while
the latter corresponds to these projections being measured with respect to
the arbitrary vector $\widehat{{\bf a}}$ whose polar angles are $(\theta
,\varphi )$. To define the generalized probability amplitude corresponding
to the latter case, we add the superscript $\widehat{{\bf a}}$ to $M$ . The
expansion for the generalized probability amplitude is thus 
\begin{equation}
\Psi (j,M^{(\widehat{{\bf a}})};\theta _1,\varphi _1,\theta _2,\varphi
_2)=\sum_{m_1,m_2}\chi (j,M^{(\widehat{{\bf a}})};m_1,m_2)\Phi
(m_1,m_2;\theta _1,\varphi _1,\theta _2,\varphi _2).  \label{fo14}
\end{equation}
In this expansion, the projections of the angular momenta of the subsystems
are not necessarily measured with respect to the direction $\widehat{{\bf a}}
$. In fact the projection of subsystem $1$ need not be measured with respect
to the same vector as the projection of subsystem $2$. In general, the
projection of the angular momentum of subsystem $1$ is measured relative to
the direction $\widehat{{\bf g}}_1$ and that of subsystem $2$ with respect
to the direction $\widehat{{\bf g}}_1$. For this reason, the condition that $%
\chi (j,M^{(\widehat{{\bf a}})};m_1,m_2)$ vanishes unless $m_1+m_2=M$ is
generally not satisfied. Furthermore, the summation is generally a double
summation, running over indices corresponding to the spin projections of
subsystems $1$ and $2$ respectively. If the projections of subsystem $1$
with respect to $\widehat{{\bf g}}_1$ are $(m_1)_\alpha ^{(\widehat{{\bf g}}%
_1)}$, while those of subsystem $2$ with respect to $\widehat{{\bf g}}_2$
are $(m_2)_{\alpha ^{^{\prime }}}^{(\widehat{{\bf g}}_2)}$, Eq. (\ref{fo14})
is modified to

\begin{eqnarray}
&&\Psi (j,M^{(\widehat{{\bf a}})};\theta _1,\varphi _1,\theta _2,\varphi
_2)=\sum_{\alpha ,\alpha ^{\prime }}\chi (j,M^{(\widehat{{\bf a}}%
)};(m_1)_\alpha ^{(\widehat{{\bf g}}_1)},(m_2)_{\alpha ^{^{\prime }}}^{(%
\widehat{{\bf g}}_2)})  \nonumber \\
&&\times \Phi ((m_1)_\alpha ^{(\widehat{{\bf g}}_1)},(m_2)_{\alpha
^{^{\prime }}}^{(\widehat{{\bf g}}_2)};\theta _1,\varphi _1,\theta
_2,\varphi _2).  \label{fo14g}
\end{eqnarray}
But in the special event that $\widehat{{\bf g}}_1=\widehat{{\bf g}}_2=%
\widehat{{\bf a}}$, then the functions $\chi (j,M^{(\widehat{{\bf a}}%
)};(m_1)_\alpha ^{(\widehat{{\bf a}})},(m_2)_{\alpha ^{^{\prime }}}^{(%
\widehat{{\bf a}})})$ are generalized Clebsch-Gordan coefficients[1] and it
is once more true that 
\begin{equation}
(m_1)_\alpha ^{(\widehat{{\bf a}})}+(m_2)_{\alpha ^{^{\prime }}}^{(\widehat{%
{\bf a}})})=M^{(\widehat{{\bf a}})}.  \label{fo14j}
\end{equation}

\subsection{Theory for Spin}

We now consider the case of spin. For a measurement on a simple system, the
initial state is defined by a spin projection with respect to a given
initial direction, while the final state is defined by a spin projection
with respect to a new final direction. Thus, in the theory outlined in the
previous section, probability amplitudes corresponding to spin projection
measurements replace the spherical harmonics.

For a system of compounded spin, the total spin is 
\begin{equation}
{\bf S}={\bf S}_1+{\bf S}_2.  \label{fi15}
\end{equation}
The quantum numbers of the spins are $s$, $s_1$ and $s_2$ for the total
system, subsystem $1$ and subsystem $2$, respectively.

Suppose that the projections of the combined spin are initially known with
respect to the direction of the vector $\widehat{{\bf a}}$, whose polar
angles are $(\theta ,\varphi )$. Let the projection of the total spin in
that direction be $M_i^{(\widehat{{\bf a}})}\hbar $. We proceed to measure
the projection of the spin of subsystem $1$ with respect to the direction $%
\widehat{{\bf c}}_1$ (defined by the angles $(\theta _1,\varphi _1)$) and
the projection of the spin of subsystem $2$ with respect to the direction $%
\widehat{{\bf c}}_2$ (polar angles $(\theta _2,\varphi _2)$). The
projections that result from the measurement are identified by their
corresponding quantum numbers and the vectors with respect to which they are
measured. Thus, the probability amplitude for this measurement is $\Psi
(s,M_i^{(\widehat{{\bf a}})};(m_1)_u^{(\widehat{{\bf c}}_1)},(m_2)_v^{(%
\widehat{{\bf c}}_2)})$, where $(u,v=1,2,...).$ The generalized probability
amplitude Eq. (\ref{fo14g}) is expressed as

\begin{eqnarray}
&&\Psi (j,M_i^{(\widehat{{\bf a}})};(m_1)_u^{(\widehat{{\bf c}}%
_1)},(m_2)_v^{(\widehat{{\bf c}}_2)})=\sum_{\alpha ,\alpha ^{^{\prime
}}}\chi (s,M_i^{(\widehat{{\bf a}})};(m_1)_\alpha ^{(\widehat{{\bf g}}%
_1)},(m_2)_{\alpha ^{^{\prime }}}^{(\widehat{{\bf g}}_2)})  \nonumber \\
&&\times \Phi ((m_1)_\alpha ^{(\widehat{{\bf g}}_1)},(m_2)_{\alpha
^{^{\prime }}}^{(\widehat{{\bf g}}_2)};(m_1)_u^{(\widehat{{\bf c}}%
_1)},(m_2)_v^{(\widehat{{\bf c}}_2)}).  \label{fi15a}
\end{eqnarray}

Since the direction vectors $\widehat{{\bf g}}_1$ and $\widehat{{\bf g}}_2$
are arbitrary, they may be chosen for best convenience. The obvious choice
is $\widehat{{\bf g}}_1=\widehat{{\bf g}}_2=\widehat{{\bf k}}$. We observe
that if this is the case, and in addition $\widehat{{\bf a}}=\widehat{{\bf k}%
}$, then the $\chi $'s become Clebsch-Gordan coefficients. Since the
Clebsch-Gordan coefficients vanish unless $(m_1)^{(\widehat{{\bf k}}%
)}+(m_2)^{(\widehat{{\bf k}})}=M^{(\widehat{{\bf k}})}$, the double
summation effectively becomes a single summation. In fact, if $\widehat{{\bf %
g}}_1$ and $\widehat{{\bf g}}_2$ are arbitrary, then the probability
amplitudes $\chi (s,M_i^{(\widehat{{\bf a}})};(m_1)_\alpha ^{(\widehat{{\bf g%
}}_1)},(m_2)_{\alpha ^{^{\prime }}}^{(\widehat{{\bf g}}_2)})$ and $\Psi
(s,M_i^{(\widehat{{\bf a}})};(m_1)_u^{(\widehat{{\bf c}}_1)},(m_2)_v^{(%
\widehat{{\bf c}}_2)})$ are essentially identical, since they differ only in
the choice of arbitrary vectors along which the spin projections of
subsystems $1$ and $2$ are measured. In practice, it is essential to make
the choice $\widehat{{\bf g}}_1=\widehat{{\bf g}}_2=\widehat{{\bf k}}$, in
order to convert $\chi (s,M_i^{(\widehat{{\bf a}})};(m_1)_\alpha ^{(\widehat{%
{\bf g}}_1)},(m_2)_{\alpha ^{^{\prime }}}^{(\widehat{{\bf g}}_2)})$ to $\chi
(s,M_i^{(\widehat{{\bf a}})};(m_1)_\alpha ^{(\widehat{{\bf k}}%
)},(m_2)_{\alpha ^{\prime }}^{(\widehat{{\bf k}})})$ which can be expressed
in terms of Clebsch-Gordan coefficients [1]. By this means it is possible to
find an expression for $\Psi (s,M_i^{(\widehat{{\bf a}})};(m_1)_u^{(\widehat{%
{\bf c}}_1)},(m_2)_v^{(\widehat{{\bf c}}_2)})$ (or equivalently $\chi
(s,M_i^{(\widehat{{\bf a}})};(m_1)_\alpha ^{(\widehat{{\bf g}}%
_1)},(m_2)_{\alpha ^{^{\prime }}}^{(\widehat{{\bf g}}_2)})$).

The actual form of the generalized spin probability amplitudes has been
obtained in Ref. [1]. It is

\begin{eqnarray}
&&\Psi (s,M_i^{(\widehat{{\bf a}})};(m_1)_u^{(\widehat{{\bf c}}%
_1)},(m_2)_v^{(\widehat{{\bf c}}_2)})=  \label{si16} \\
&&\sum_{\alpha ,\alpha ^{\prime }}\sum_l\zeta (s,M_i^{(\widehat{{\bf a}}%
)};s,M_l^{(\widehat{{\bf k}})})\vartheta (s,M_l^{(\widehat{{\bf k}}%
)};(m_1)_\alpha ^{(\widehat{{\bf k}})},(m_2)_{\alpha ^{\prime }}^{(\widehat{%
{\bf k}})})  \nonumber \\
&&\times \Phi ((m_1)_\alpha ^{(\widehat{{\bf k}})},(m_2)_{\alpha ^{\prime
}}^{(\widehat{{\bf k}})};(m_1)_u^{(\widehat{{\bf c}}_1)},(m_2)_v^{(\widehat{%
{\bf c}}_2)}).  \nonumber
\end{eqnarray}

The quantities in expression (\ref{si16}) are defined as follows. The
quantity $\zeta (s,M_i^{(\widehat{{\bf a}})};s,M_l^{(\widehat{{\bf k}})})$
is the probability amplitude that if the total spin is $s$ and its
projection along the vector $\widehat{{\bf a}}$ is $M_i^{(\widehat{{\bf a}}%
)}\hbar $, a measurement of its projection along the $z$ axis gives $M_l^{(%
\widehat{{\bf k}})}\hbar .$ The quantity $\vartheta (s,M_l^{(\widehat{{\bf k}%
})};(m_1)_\alpha ^{(\widehat{{\bf k}})},(m_2)_{\alpha ^{\prime }}^{(\widehat{%
{\bf k}})})$ is actually the standard Clebsch-Gordan coefficient for the
case at hand. It is obtained from $\chi (s,M_i^{(\widehat{{\bf a}}%
)};(m_1)_\alpha ^{(\widehat{{\bf g}}_1)},(m_2)_{\alpha ^{^{\prime }}}^{(%
\widehat{{\bf g}}_2)})$ by setting $\widehat{{\bf a}}=\widehat{{\bf g}}_1=%
\widehat{{\bf g}}_2=\widehat{{\bf k}}$.

For future convenience we rewrite Eq. (\ref{si16}) as 
\begin{eqnarray}
&&\Psi (s,M_i^{(\widehat{{\bf a}})}{};(m_1)_u^{(\widehat{{\bf c}}%
_1)},(m_2)_v^{(\widehat{{\bf c}}_2)})=\sum_{\alpha ,\alpha ^{\prime }}\chi
(s,M_i^{(\widehat{{\bf a}})};(m_1)_\alpha ^{(\widehat{{\bf k}}%
)},(m_2)_{\alpha ^{\prime }}^{(\widehat{{\bf k}})})  \nonumber \\
&&\times \Phi ((m_1)_\alpha ^{(\widehat{{\bf k}})},(m_2)_{\alpha ^{\prime
}}^{(\widehat{{\bf k}})};(m_1)_u^{(\widehat{{\bf c}}_1)},(m_2)_v^{(\widehat{%
{\bf c}}_2)}),  \label{se17}
\end{eqnarray}
where

\begin{eqnarray}
\ &&\chi (s,M_i^{(\widehat{{\bf a}})};(m_1)_\alpha ^{(\widehat{{\bf k}}%
)},(m_2)_{\alpha ^{\prime }}^{(\widehat{{\bf k}})})=\sum_l\zeta (s,M_i^{(%
\widehat{{\bf a}})};s,M_l^{(\widehat{{\bf k}})})  \nonumber \\
\ &&\times \vartheta (s,M_l^{(\widehat{{\bf k}})};(m_1)_\alpha ^{(\widehat{%
{\bf k}})},(m_2)_{\alpha ^{\prime }}^{(\widehat{{\bf k}})}).  \label{ei18}
\end{eqnarray}

If we compare Eq. (\ref{se17}) with the fundamental expansion Eq. (\ref{tw2}%
), we see that the intermediate observable which we are using to achieve the
expansion is the combination of spin projections of systems $1$ and $2$ with
respect to the $z$ axis. The notation is simpler if we use the symbol $B$
for this observable. However, because of the double summation, the symbol
has two subscripts. If the subsystems $1$ and $2$ are both spin-$1/2$
systems, the values of $B$ are 
\begin{equation}
B_{11}=((m_1)_1^{(\widehat{{\bf k}})},(m_2)_1^{({\bf k})})=((+\frac 12)^{(%
\widehat{{\bf k}})},(+\frac 12)^{({\bf k})}),  \label{ni19}
\end{equation}
\begin{equation}
B_{12}=((m_1)_1^{(\widehat{{\bf k}})},(m_2)_2^{({\bf k})})=((+\frac 12)^{(%
\widehat{{\bf k}})},(-\frac 12)^{({\bf k})}),  \label{tw20}
\end{equation}
\begin{equation}
B_{21}=((m_1)_2^{(\widehat{{\bf k}})},(m_2)_1^{({\bf k})})=((-\frac 12)^{(%
\widehat{{\bf k}})},(+\frac 12)^{({\bf k})})  \label{tw21}
\end{equation}
and 
\begin{equation}
B_{22}=((m_1)_2^{(\widehat{{\bf k}})},(m_2)_2^{({\bf k})})=((-\frac 12)^{(%
\widehat{{\bf k}})},(-\frac 12)^{({\bf k})}).  \label{tw22}
\end{equation}

This makes it easier to denote values pertaining to just one subsystem. For
a particular value $B_{\alpha \alpha ^{\prime }}$, we shall denote the value
corresponding to the subsystem $w$ by $(B_{\alpha \alpha ^{\prime }})_w$,$\;$%
where $w=1,2$. This means that for subsystem $1$, 
\begin{equation}
(B_{11})_1=(B_{12})_1=(+\frac 12)^{(\widehat{{\bf k}})}  \label{tw23}
\end{equation}
and 
\begin{equation}
(B_{21})_1=(B_{22})_1=(-\frac 12)^{(\widehat{{\bf k}})},  \label{tw24}
\end{equation}
while for subsystem $2$, 
\begin{equation}
(B_{11})_2=(B_{21})_2=(+\frac 12)^{(\widehat{{\bf k}})}  \label{tw25}
\end{equation}
and 
\begin{equation}
(B_{12})_2=(B_{22})_2=(-\frac 12)^{(\widehat{{\bf k}})}.  \label{tw26}
\end{equation}

For the sake of convenience, we set $A_i=(s,M_i^{(\widehat{{\bf a}})})$.
Then the probability amplitude Eq. (\ref{se17}) becomes

\begin{equation}
\Psi (A_i;(m_1)_u^{(\widehat{{\bf c}}_1)},(m_2)_v^{(\widehat{{\bf c}}%
_2)})=\sum_{\alpha ,\alpha ^{\prime }}\chi (A_i;B_{\alpha \alpha ^{\prime
}})\Phi (B_{\alpha \alpha ^{\prime }};(m_1)_u^{(\widehat{{\bf c}}%
_1)},(m_2)_v^{(\widehat{{\bf c}}_2)}).  \label{tw27}
\end{equation}
We remind ourselves that according to Eq. (\ref{th13b}), 
\begin{eqnarray}
&&\Phi (B_{\alpha \alpha ^{\prime }};(m_1)_u^{(\widehat{{\bf c}}%
_1)},(m_2)_v^{(\widehat{{\bf c}}_2)})=\Phi ((m_1)_\alpha ^{(\widehat{{\bf k}}%
)},(m_2)_{\alpha ^{\prime }}^{(\widehat{{\bf k}})};(m_1)_u^{(\widehat{{\bf c}%
}_1)},(m_2)_v^{(\widehat{{\bf c}}_2)})  \nonumber \\
\ &=&\phi _1((m_1)_\alpha ^{(\widehat{{\bf k}})};(m_1)_u^{(\widehat{{\bf c}}%
_1)})\phi _2((m_2)_{\alpha ^{\prime }}^{({\bf k})}:(m_2)_v^{(\widehat{{\bf c}%
}_2)}).  \label{tw28}
\end{eqnarray}
We remark that, purely for convenience, we have altered the notation
slightly. Thus, $\phi _i=\phi ^{(i)}$ $(i=1,2)$.

\section{From Probability-Amplitude Mechanics to Matrix Mechanics}

\subsection{Theory For Simple Systems}

In order to move from wave to matrix mechanics, an expansion of the
eigenfunction or wave function in terms of some basis set is necessary. To
obtain the matrix theory of orbital angular momentum we use the spherical
harmonics as the basis set. This kind of procedure was thought impossible
for spin, because spin is not describable by eigenfunctions resulting from
an eigenvalue equation. But in our work[1, 7-11], we have shown how, by
using the probability amplitudes for measurements on spin systems, we can
derive the matrix treatment of spin in the same way as the matrix treatment
of orbital angular momentum is obtained.

The relation Eq. (\ref{tw2}) is the basis of the transformation from
amplitude to matrix mechanics. We now review how we use it to obtain the
matrix treatment of a simple quantum system. This review is needed because
it is the foundation of the more involved derivation of the standard matrix
treatment of compounded spin from generalized probability amplitudes.

Let us suppose that have a quantum system possessing the observables $A$, $B$
and $C$. We assume that as we are measuring values of $C$, we are measuring
values of a quantity $T(C)$ which is a function of $C$. Let $T(C)$ take upon
measurement the values $T_n$ determined by the values $C_n$ of $C$. Thus $%
T_n=T(C_n)$. The expectation value of $T$ is

\begin{equation}
\left\langle T(C)\right\rangle =\sum_{n=1}^N\left| \psi (A_i;C_n)\right|
^2T_n.  \label{tw28a}
\end{equation}
where $N$ is the total number of eigenvalues of $C$.

If we use the expansions 
\begin{equation}
\psi ^{*}(A_i;C_n)=\sum_{j=1}^N\eta ^{*}(A_i;B_j)\xi ^{*}(B_j;C_n)
\label{fo4}
\end{equation}
and 
\begin{equation}
\psi (A_i;C_n)=\sum_{j^{\prime }=1}^N\eta (A_i;B_{j^{\prime }})\xi
(B_{j^{\prime }};C_n),  \label{fi5}
\end{equation}
we find 
\begin{equation}
\left\langle T(C)\right\rangle =\sum_{j=1}^N\sum_{j^{\prime }=1}^N\eta
^{*}(A_i;B_j)T_{jj^{\prime }}\eta (A_i;B_{j^{\prime }}),  \label{si6}
\end{equation}
where 
\begin{equation}
T_{jj^{\prime }}=\sum_{n=1}^N\xi ^{*}(B_j;C_n)T_n\xi (B_{j^{\prime }};C_n).
\label{se7}
\end{equation}

Hence 
\begin{equation}
\left\langle T(C)\right\rangle =[\eta (A_i)]^{\dagger }[T][\eta (A_i)],
\label{ei8}
\end{equation}
where the state is

\begin{equation}
\lbrack \eta (A_i)]=\left( 
\begin{array}{c}
\eta (A_i;B_1) \\ 
\eta (A_i;B_2) \\ 
.. \\ 
\eta (A_i;B_N)
\end{array}
\right) ,  \label{ni9}
\end{equation}
and the operator is

\begin{equation}
\lbrack T]=\left( 
\begin{array}{cccc}
T_{11} & T_{12} & ... & T_{1N} \\ 
T_{21} & T_{22} & ... & T_{2N} \\ 
... & ... & ... & ... \\ 
T_{N1} & T_{N2} & ... & T_{NN}
\end{array}
\right) .  \label{te10}
\end{equation}

Thus, by means of the probability amplitudes for a quantum system, we can
derive its matrix treatment. We note the convention of enclosing a quantity
in brackets in order to denote its matrix representation.

\subsection{Theory for Systems of Compounded Spin}

The theory in the previous section will now be extended so as to yield the
derivation of the matrix treatment of systems of compounded spin. As usual,
we go through the expectation value in order to obtain the matrix form of
the probability amplitudes for compounded spin, and of the operators for
quantities that may be measured on the systems.

Suppose that a compounded spin is obtained by adding the spins $s_1$ and $%
s_2 $. Suppose that initially the total spin is $s$ and its projection with
respect to the vector $\widehat{{\bf a}}$ is $M\hbar $. Subsequently, the
spin projection of the spin of system $1$ is measured along the vector $%
\widehat{{\bf c}}_1$ and the spin projection of the spin of system $2$ is
measured along the vector $\widehat{{\bf c}}_2$. At the same time, the
quantity $R((m_1)^{(\widehat{{\bf c}}_1)},(m_2)^{(\widehat{{\bf c}}_2)})$ is
measured. This quantity is measured on the separate systems $1$ and $2$. It
is constructed from the quantity $r^{(1)}((m_1)^{(\widehat{{\bf c}}_1)})$,
which is measured on system $1$ and the quantity $r^{(2)}((m_2)^{(\widehat{%
{\bf c}}_2)})$ measured on system $2$. Therefore, we write 
\begin{equation}
R=R(r^{(1)}((m_1)^{(\widehat{{\bf c}}_1)}),r^{(2)}((m_2)^{(\widehat{{\bf c}}%
_2)}).  \label{tw28x}
\end{equation}
The values of $r^{(1)}((m_1)^{(\widehat{{\bf c}}_1)})$ are independent of
the values of $r^{(2)}((m_2)^{(\widehat{{\bf c}}_2)})$: any value $%
r^{(1)}((m_1)_u^{(\widehat{{\bf c}}_1)})$ and any value $r^{(2)}((m_2)_v^{(%
\widehat{{\bf c}}_2)})$ can result together from the measurements.
Therefore, the expectation value of $R$ is

\begin{eqnarray}
&&\left\langle R\right\rangle =\sum_u\sum_v\Psi ^{*}(A_i;(m_1)_u^{(\widehat{%
{\bf c}}_1)},(m_2)_v^{(\widehat{{\bf c}}_2)})\Psi (A_i;(m_1)_u^{(\widehat{%
{\bf c}}_1)},(m_2)_v^{(\widehat{{\bf c}}_2)})  \nonumber \\
&&\times R(r^{(1)}((m_1)_u^{(\widehat{{\bf c}}_1)}),r^{(2)}((m_2)_v^{(%
\widehat{{\bf c}}_2)}),  \label{tw29}
\end{eqnarray}
where $R(r^{(1)}((m_1)_u^{(\widehat{{\bf c}}_1)}),r^{(2)}((m_2)_v^{(\widehat{%
{\bf c}}_2)})$ is an actual value of $R((m_1)^{(\widehat{{\bf c}}%
_1)},(m_2)^{(\widehat{{\bf c}}_2)}).$

The probability amplitude is given by Eq.(\ref{tw27}). Using Eq. (\ref{tw28}%
), we obtain the expansions 
\begin{eqnarray}
&&\Psi ^{*}(A_i;(m_1)_u^{(\widehat{{\bf c}}_1)},(m_2)_v^{(\widehat{{\bf c}}%
_2)})=\sum_{\alpha ,\alpha ^{\prime }}\chi ^{*}(A_i;B_{\alpha \alpha
^{\prime }})  \nonumber \\
\ &&\times \phi _1^{*}((B_{\alpha \alpha ^{\prime }})_1;(m_1)_u^{(\widehat{%
{\bf c}}_1)})\phi _2^{*}((B_{\alpha \alpha ^{\prime }})_2;(m_2)_v^{(\widehat{%
{\bf c}}_2)})  \label{th30}
\end{eqnarray}
and 
\begin{eqnarray}
&&\Psi (A_i;(m_1)_u^{(\widehat{{\bf c}}_1)},(m_2)_v^{(\widehat{{\bf c}}%
_2)})=\sum_{\beta ,\beta ^{\prime }}\chi (A_i;B_{\beta \beta ^{\prime }}) 
\nonumber \\
&&\times \phi _1((B_{\beta \beta ^{\prime }})_1;(m_1)_u^{(\widehat{{\bf c}}%
_1)})\phi _2((B_{\beta \beta ^{\prime }})_2;(m_2)_v^{(\widehat{{\bf c}}_2)}).
\label{th31}
\end{eqnarray}
The expectation value becomes

\begin{eqnarray}
\left\langle R\right\rangle &=&\sum_{\alpha ,\alpha ^{\prime }}\sum_{\beta
,\beta ^{\prime }}\chi ^{*}(A_i;B_{\alpha \alpha ^{\prime }})\chi
(A_i;B_{\beta \beta ^{\prime }})  \nonumber \\
&&\times \sum_u\sum_v\{\phi _1^{*}((B_{\alpha \alpha ^{\prime
}})_1;(m_1)_u^{(\widehat{{\bf c}}_1)})\phi _1((B_{\beta \beta ^{\prime
}})_1;(m_1)_u^{(\widehat{{\bf c}}_1)})  \nonumber \\
\ &&\times \phi _2^{*}((B_{\alpha \alpha ^{\prime }})_2;(m_2)_v^{(\widehat{%
{\bf c}}_2)})\phi _2((B_{\beta \beta ^{\prime }})_2;(m_2)_v^{(\widehat{{\bf c%
}}_2)})  \nonumber \\
&&\times R(r^{(1)}((m_1)_u^{(\widehat{{\bf c}}_1)})r^{(2)}((m_2)_v^{(%
\widehat{{\bf c}}_2)})\}.  \label{th32}
\end{eqnarray}

When $R$ is factorizable, so that 
\begin{equation}
R(r^{(1)}((m_1)^{(\widehat{{\bf c}}_1)}),r^{(2)}((m_2)^{(\widehat{{\bf c}}%
_2)})=r^{(1)}((m_1)^{(\widehat{{\bf c}}_1)})r^{(2)}((m_2)^{(\widehat{{\bf c}}%
_2)}),  \label{th32b}
\end{equation}
we can write Eq. (\ref{th32}) as 
\begin{eqnarray}
&&\left\langle R\right\rangle =\sum_\alpha \sum_{\alpha ^{^{\prime }}}\chi
^{*}(A_i;B_{\alpha \alpha ^{\prime }})\chi (A_i;B_{\beta \beta ^{\prime
}})\sum_u\phi _1^{*}((B_{\alpha \alpha ^{\prime }})_1;(m_1)_u^{(\widehat{%
{\bf c}}_1)})  \nonumber \\
\ &&\times \phi _1((B_{\beta \beta ^{\prime }})_1;(m_1)_u^{(\widehat{{\bf c}}%
_1)})r^{(1)}((m_1)_u^{(\widehat{{\bf c}}_1)})  \nonumber \\
\ &&\times \sum\limits_v\phi _2^{*}((B_{\alpha \alpha ^{\prime
}})_2;(m_2)_v^{(\widehat{{\bf c}}_2)})\phi _2((B_{\beta \beta ^{\prime
}})_2;(m_2)_v^{(\widehat{{\bf c}}_2)})r^{(2)}((m_2)_v^{(\widehat{{\bf c}}%
_2)}).  \label{th32a}
\end{eqnarray}

Thus

\begin{equation}
\left\langle R\right\rangle \ =\sum_{\alpha ,\alpha ^{\prime }}\sum_{\beta
,\beta ^{\prime }}\chi ^{*}(A_i;B_{\alpha \alpha ^{\prime }})I_{\alpha
\alpha ^{\prime }\beta \beta ^{\prime }}^{(1)}I_{\alpha \alpha ^{\prime
}\beta \beta ^{\prime }}^{(2)}\chi (A_i;B_{\beta \beta ^{\prime }}),
\label{th32e}
\end{equation}
where

\begin{eqnarray}
I_{\alpha \alpha ^{\prime }\beta \beta ^{\prime }}^{(1)} &=&\sum_u\phi
_1^{*}((B_{\alpha \alpha ^{\prime }})_1;(m_1)_u^{(\widehat{{\bf c}}%
_1)})r^{(1)}((m_1)_u^{(\widehat{{\bf c}}_1)})  \nonumber \\
&&\times \phi _1((B_{\beta \beta ^{\prime }})_1;(m_1)_u^{(\widehat{{\bf c}}%
_1)})  \label{th33}
\end{eqnarray}
and

\begin{eqnarray}
I_{\alpha \alpha ^{\prime }\beta \beta ^{\prime }}^{(2)} &=&\sum_v\phi
_2^{*}((B_{\alpha \alpha ^{\prime }})_2;(m_2)_v^{(\widehat{{\bf c}}%
_2)})r^{(2)}((m_2)_v^{(\widehat{{\bf c}}_2)})  \nonumber \\
&&\times \phi _2((B_{\beta \beta ^{\prime }})_2;(m_2)_v^{(\widehat{{\bf c}}%
_2)}).  \label{th34}
\end{eqnarray}

In order to treat $I_{\alpha \alpha ^{\prime }\beta \beta ^{\prime }}^{(1)}$%
, we introduce the observable $D$ which corresponds to spin projections of
subsystem $1$ with respect to the direction $\widehat{{\bf d}\text{,}}$
whose polar angles are $(\theta _d,\varphi _d)$. We use this observable to
expand $\phi _1$ by means of formula (\ref{tw2}). We note that the values $%
D_p$ are

\begin{equation}
D_p=(m_1)_p^{(\widehat{{\bf d}})}\hbar .  \label{th34a}
\end{equation}
The expansions of $\phi _1$ and $\phi _1^{*}$ are 
\begin{equation}
\phi _1^{*}((B_{\alpha \alpha ^{\prime }})_1;(m_1)_u^{(\widehat{{\bf c}}%
_1)})=\sum_p\eta _1^{*}((B_{\alpha \alpha ^{\prime }})_1;D_p)\xi
_1^{*}(D_p;(m_1)_u^{(\widehat{{\bf c}}_1)})  \label{th35}
\end{equation}
and

\begin{equation}
\phi _1((B_{\beta \beta ^{\prime }})_1;(m_1)_u^{(\widehat{{\bf c}}%
_1)})=\sum_{p^{^{\prime }}}\eta _1((B_{\beta \beta ^{\prime
}})_1;D_{p^{^{\prime }}})\xi _1(D_{p^{^{\prime }}};(m_1)_u^{(\widehat{{\bf c}%
}_1)}).  \label{th36}
\end{equation}
Applying the theory outlined in Section $4.1$, we obtain

\begin{equation}
I_{\alpha \alpha ^{\prime }\beta \beta ^{\prime
}}^{(1)}=\sum_p\sum_{p^{^{\prime }}}\eta _1^{*}((B_{\alpha \alpha ^{\prime
}})_1;D_p)r_{pp^{^{\prime }}}^{(1)}\eta _1((B_{\beta \beta ^{\prime
}})_1;D_{p^{^{\prime }}}),  \label{th37}
\end{equation}
where 
\begin{equation}
r_{pp^{^{\prime }}}^{(1)}=\sum_u\xi _1^{*}(D_p;(m_1)_u^{(\widehat{{\bf c}}%
_1)})r^{(1)}((m_1)_u^{(\widehat{{\bf c}}_1)})\xi _1(D_{p^{^{\prime
}}};(m_1)_u^{(\widehat{{\bf c}}_1)}).  \label{th38}
\end{equation}

Hence

\begin{equation}
I_{\alpha \alpha ^{\prime }\beta \beta ^{\prime }}^{(1)}=[\eta _1((B_{\alpha
\alpha ^{\prime }})_1)]^{\dagger }[r^{(1)}][\eta _1((B_{\beta \beta ^{\prime
}})_1)],  \label{th39}
\end{equation}
where 
\begin{equation}
\lbrack \eta _1(B_{\alpha \alpha ^{\prime }})]=\left( 
\begin{array}{c}
\eta _1((B_{\alpha \alpha ^{\prime }})_1;D_1) \\ 
\eta _1((B_{\alpha \alpha ^{\prime }})_1;D_2) \\ 
: \\ 
\eta _1((B_{\alpha \alpha ^{\prime }})_1;D_N)
\end{array}
\right)  \label{fo40}
\end{equation}
and 
\begin{equation}
\lbrack r^{(1)}]=\left( 
\begin{array}{cccc}
r_{11}^{(1)} & r_{12}^{(1)} & .. & r_{1N}^{(1)} \\ 
r_{21}^{(1)} & r_{22}^{(1)} & .. & r_{2N}^{(1)} \\ 
.. & .. & .. & .. \\ 
r_{N1}^{(1)} & r_{N2}^{(1)} & .. & r_{NN}^{(1)}
\end{array}
\right) .  \label{fo41}
\end{equation}
Here $N$ is the total number of states of the observable $D$. In the case
where subsystem $1$ is a spin-$1/2$ system, $N=2$.

To deal with $I_{\alpha \alpha ^{\prime }\beta \beta ^{\prime }}^{(2)}$, we
introduce the observable $F$ defined by the unit vector $\widehat{{\bf f}}$
whose polar angles are $(\theta _f,\varphi _f)$. The eigenvalues of $F$ are 
\begin{equation}
F_q=(m_2)_q^{(\widehat{{\bf f}})}\hbar .  \label{fo44x}
\end{equation}
Expanding $\phi _2$ over the states of $F$, we get 
\begin{equation}
\phi _2^{*}(B_{\alpha \alpha ^{\prime }};(m_2)_v^{(\widehat{{\bf c}}%
_2)})=\sum_q\eta _2^{*}((B_{\alpha \alpha ^{\prime }})_2;F_q)\xi
_2^{*}(F_q;(m_2)_v^{(\widehat{{\bf c}}_2)})  \label{fo41c}
\end{equation}
and

\begin{equation}
\phi _2(B_{\beta \beta ^{\prime }};(m_2)_v^{(\widehat{{\bf c}}%
_2)})=\sum_{q^{^{\prime }}}\eta _2((B_{\beta \beta ^{\prime
}})_2;F_{q^{^{\prime }}})\xi _2(F_{q^{^{\prime }}};(m_2)_v^{(\widehat{{\bf c}%
}_2)}).  \label{fo42}
\end{equation}
This means that

\begin{equation}
I_{\alpha \alpha ^{\prime }\beta \beta ^{\prime
}}^{(2)}=\sum_q\sum_{q^{^{\prime }}}\eta _2^{*}((B_{\alpha \alpha ^{\prime
}})_2;F_q)r_{qq^{^{\prime }}}^{(2)}\eta _2((B_{\beta \beta ^{\prime
}})_2;F_{q^{^{\prime }}}),  \label{fo43}
\end{equation}
where 
\begin{equation}
r_{qq^{^{\prime }}}^{(2)}=\sum_v\xi _2^{*}(F_q;(m_2)_v^{(\widehat{{\bf c}}%
_2)})r^{(2)}((m_2)_v^{(\widehat{{\bf c}}_2)})\xi _2(F_{q^{^{\prime
}}};(m_2)_v^{(\widehat{{\bf c}}_2)}).  \label{fo44}
\end{equation}

In view of Eq. (\ref{fo43}), the expression for $I_{\alpha \alpha ^{\prime
}\beta \beta ^{\prime }}^{(2)}$ is 
\begin{equation}
I_{\alpha \alpha ^{\prime }\beta \beta ^{\prime }}^{(2)}=[\eta _2((B_{\alpha
\alpha ^{\prime }})_2)]^{\dagger }[r^{(2)}][\eta _2((B_{\beta \beta ^{\prime
}})_2)],  \label{fo45}
\end{equation}
where 
\begin{equation}
\lbrack \eta _2((B_{\alpha \alpha ^{\prime }})_2)]=\left( 
\begin{array}{c}
\eta _2((B_{\alpha \alpha ^{\prime }})_2;F_1) \\ 
\eta _2((B_{\alpha \alpha ^{\prime }})_2;F_2) \\ 
: \\ 
\eta _2((B_{\alpha \alpha ^{\prime }})_2;F_M)
\end{array}
\right)  \label{fo46}
\end{equation}
and 
\begin{equation}
\lbrack r^{(2)}]=\left( 
\begin{array}{cccc}
r_{11}^{(2)} & r_{12}^{(2)} & .. & r_{1M}^{(2)} \\ 
r_{21}^{(2)} & r_{22}^{(2)} & .. & r_{2M}^{(2)} \\ 
.. & .. & .. & .. \\ 
r_{M1}^{(2)} & r_{M2}^{(2)} & .. & r_{MM}^{(2)}
\end{array}
\right) .  \label{fo47}
\end{equation}
Here $M$ is the number of states of $F$.

Collecting all the results together, we find that

\begin{eqnarray}
&&{}\left\langle R\right\rangle =\sum_{\alpha ,\alpha ^{\prime }}\sum_{\beta
,\beta ^{\prime }}\chi ^{*}(A_i;B_{\alpha \alpha ^{\prime }})[\eta
_1((B_{\alpha \alpha ^{\prime }})_1)]^{\dagger }[r^{(1)}][\eta _1((B_{\beta
\beta ^{\prime }})_1)]  \nonumber \\
&&\times [\eta _2((B_{\alpha \alpha ^{\prime }})_2)]^{\dagger
}[r^{(2)}][\eta _2((B_{\beta \beta ^{\prime }})_2)]\chi (A_i;B_{\beta \beta
^{\prime }})  \nonumber \\
\ &=&{}\left( \sum_{\alpha ,\alpha ^{\prime }}\chi ^{*}(A_i;B_{\alpha \alpha
^{\prime }})[\eta _1((B_{\alpha \alpha ^{\prime }})_1)]^{\dagger }[\eta
_2((B_{\alpha \alpha ^{\prime }})_2)]^{\dagger }\right) [r^{(1)}][r^{(2)}] 
\nonumber \\
&&\times \left( \sum_{\beta ,\beta ^{\prime }}\chi (A_i;B_{\beta \beta
^{\prime }})[\eta _1((B_{\beta \beta ^{\prime }})_1)][\eta _2((B_{\beta
\beta ^{\prime }})_2)]\right)  \nonumber \\
\ &=&{}[\Psi (A_i;(m_1)^{(\widehat{{\bf c}}_1)},(m_2)^{(\widehat{{\bf c}}%
_2)})]^{\dagger }[R(r^{(1)}((m_1)^{(\widehat{{\bf c}}_1)}),r^{(2)}((m_2)^{(%
\widehat{{\bf c}}_2)}))]  \nonumber \\
&&{}\times [\Psi (A_i;(m_1)^{(\widehat{{\bf c}}_1)},(m_2)^{(\widehat{{\bf c}}%
_2)})],  \label{fo48}
\end{eqnarray}
where

\begin{equation}
\lbrack \Psi (A_i;(m_1)^{(\widehat{{\bf c}}_1)},(m_2)^{(\widehat{{\bf c}}%
_2)})]=\left( \sum_{\alpha ,\alpha ^{\prime }}\chi (A_i;B_{\alpha \alpha
^{\prime }})[\eta _1((B_{\alpha \alpha ^{\prime }})_1)][\eta _2((B_{\alpha
\alpha ^{\prime }})_2)]\right)  \label{fo49}
\end{equation}
and

\begin{equation}
\lbrack R(r^{(1)}((m_1)^{\widehat{{\bf c}}_1}),r^{(2)}((m_2)^{(\widehat{{\bf %
c}}_2)}))]=[r^{(1)}][r^{(2)}].  \label{fi50}
\end{equation}

We see that in Eq. (\ref{fo49}) we have obtained the generalized vector
state corresponding to the probability amplitude $\Psi (A_i;(m_1)_u^{(%
\widehat{{\bf c}}_1)},(m_2)_v^{(\widehat{{\bf c}}_2)}).$ In addition, in
Eqs. (\ref{th38}), \ref{fo41}, (\ref{fo44}) and (\ref{fo47}), we have
obtained the generalized operator for any observable of the system which is
a function of the spin projections of the subsystems.

We note that by definition, the vectors relating to subsystem $1$ act only
on one another and on the operator corresponding to this subsystem. The same
holds for the quantities corresponding to subsystem $2$. To emphasize this
fact, we introduce the labels $1$ and $2$ to distinguish the corresponding
quantities. Thus, we get 
\begin{equation}
\lbrack \Psi (A_i;(m_1)^{(\widehat{{\bf c}}_1)},(m_2)^{(\widehat{{\bf c}}%
_2)}]=\left( \sum_{\alpha ,\alpha ^{\prime }}\chi (A_i;B_{\alpha \alpha
^{\prime }})[\eta _1((B_{\alpha \alpha ^{\prime }})_1)]_1[\eta _2((B_{\alpha
\alpha ^{\prime }})_2)]_2\right)   \label{fi51}
\end{equation}
and 
\begin{equation}
\lbrack R(r^{(1)}((m_1)^{(\widehat{{\bf c}}_1)}),r^{(2)}((m_2)^{(\widehat{%
{\bf c}}_2)})]=[r^{(1)}((m_1)^{(\widehat{{\bf c}}_1)})]_1[r^{(2)}((m_2)^{(%
\widehat{{\bf c}}_2)})]_2.  \label{fi52}
\end{equation}

These results are dependent on the condition that $R$ is factorizable; if
this is not the case, it is not so easy to transform Eq. (\ref{th32}) to
matrix form. But in that case, we can use the alternative approach of Ref.
[1]; we then end up with $3$- or $4$- dimensional matrix representations
which appply whether $R$ is factorizable or not.

\section{Application to Actual Systems}

\subsection{The Matrix Operator}

The results derived in the last section will now be used to obtain specific
operators and vectors. For a particular case, the operator is calculated by
explicitly working out the matrix elements $r_{pp^{^{\prime }}}^{(1)}$ and $%
r_{qq^{^{\prime }}}^{(2)}$ . The cases at hand are of a system of total spin 
$0$, and of a system of total spin $1$ (with three possible values of the
magnetic quantum number) obtained by adding two spins of $1/2$ each. The
form of the operator is independent of whether the total spin is $0$ or $1$.
We therefore derive this quantity first.

We start with the matrix element $r_{pp^{^{\prime }}}^{(1)}.$ In order to
obtain this quantity, we require the forms of the probability amplitudes $%
\xi $. Since both systems $1$ and $2$ are spin-$1/2$ systems, these are
obtained from the generalized spin-$1/2$ amplitudes, whose explicit forms we
have already worked out [7,8,10].

We first recall the details of the probability amplitudes for spin $1/2$. We
consider system $1$. Let the spin projection be initially known with respect
to ${\bf d}$; it is subsequently measured with respect to $\widehat{{\bf c}}%
_1$. The probability amplitude that it will be found upon measurement to be
up with respect to $\widehat{{\bf c}}_1$ is $\xi _1((+\frac 12)^{(\widehat{%
{\bf d}})};(+\frac 12)^{(\widehat{{\bf c}}_1)})$. The other three
probability amplitudes are therefore $\xi _1((+\frac 12)^{(\widehat{{\bf d}}%
)};(-\frac 12)^{(\widehat{{\bf c}}_1)})$, $\xi _1((-\frac 12)^{(\widehat{%
{\bf d}})};(+\frac 12)^{(\widehat{{\bf c}}_1)})$ and $\xi _1((-\frac 12)^{(%
\widehat{{\bf d}})};(-\frac 12)^{(\widehat{{\bf c}}_1)}).$ These probability
amplitudes come in a variety of forms, depending on the phase choice made
when they are being derived [10]. One form is the following:

\begin{equation}
\xi _1((+\frac 12)^{(\widehat{{\bf d}})};(+\frac 12)^{(\widehat{{\bf c}}%
_1)})=\cos \theta _d/2\cos \theta _1/2+e^{i(\varphi _d-\varphi _1)}\sin
\theta _d/2\sin \theta _1/2,  \label{fi53}
\end{equation}

\begin{equation}
\xi _1((+\frac 12)^{(\widehat{{\bf d}})};(-\frac 12)^{(\widehat{{\bf c}}%
_1)})=-\cos \theta _d/2\sin \theta _1/2+e^{i(\varphi _d-\varphi _1)}\sin
\theta _d/2\cos \theta _1/2,  \label{fi54}
\end{equation}

\begin{equation}
\xi _1((-\frac 12)^{(\widehat{{\bf d}})};(+\frac 12)^{(\widehat{{\bf c}}%
_1)})=-\sin \theta _d/2\cos \theta _1/2+e^{i(\varphi _d-\varphi _1)}\cos
\theta _d/2\sin \theta _1/2  \label{fi55}
\end{equation}
and

\begin{equation}
\xi _1((-\frac 12)^{(\widehat{{\bf d}})};(-\frac 12)^{(\widehat{{\bf c}}%
_1)})=\sin \theta _d/2\sin \theta _1/2+e^{i(\varphi _d-\varphi _1)}\cos
\theta _d/2\cos \theta _1/2.  \label{fi56}
\end{equation}

Since in this case system $2$ is also a spin-$1/2$ system, the probability
amplitudes corresponding to it are identical in form to Eqs. (\ref{fi53}) - (%
\ref{fi56}). To obtain them, we merely make the following change to the
labels: $2$ replaces subscript $1$; $\widehat{{\bf f}}$ replaces $\widehat{%
{\bf d}}$, so that $f$ replaces $d$; and $\widehat{{\bf c}}_2$ replaces $%
\widehat{{\bf c}}_1$.

For the case of spin $1/2$, the summation over $u$ which appears in the
expression for $r_{pp^{^{\prime }}}^{(1)}$ contains only two terms. $u=1$
corresponds to the outcome $(+\frac 12)^{(\widehat{{\bf c}}_1)}$ while $u=2$
corresponds to $(-\frac 12)^{(\widehat{{\bf c}}_1)}.$ Thus

\begin{eqnarray}
r_{11}^{(1)} &=&\left| \xi _1((+\frac 12)^{(\widehat{{\bf d}})};(+\frac
12)^{(\widehat{{\bf c}}_1)})\right| ^2r^{(1)}((+\frac 12)^{(\widehat{{\bf c}}%
_1)})  \nonumber \\
&&+\left| \xi _1((+\frac 12)^{(\widehat{{\bf d}})};(-\frac 12)^{(\widehat{%
{\bf c}}_1)})\right| ^2r^{(1)}((-\frac 12)^{(\widehat{{\bf c}}_1)}).
\label{fi57}
\end{eqnarray}

The values of the summation indices $p$ and $p^{^{\prime }}$are such that $%
p,p^{^{\prime }}=+1$ corresponds to $(+\frac 12)^{(\widehat{{\bf d}})}$,
while $p,p^{^{\prime }}=2$ corresponds to $(-\frac 12)^{(\widehat{{\bf d}})} 
$. Hence 
\begin{eqnarray}
\ &&r_{11}^{(1)}=[\cos ^2(\theta _d-\theta _1)/2-\sin \theta _d\sin \theta
_1\sin ^2(\varphi _d-\varphi _1)/2]r^{(1)}((+\frac 12)^{(\widehat{{\bf c}}%
_1)})  \nonumber \\
&&\ \ \ +[\sin ^2(\theta _d-\theta _1)/2+\sin \theta _d\sin \theta _1\sin
^2(\varphi _d-\varphi _1)/2]r^{(1)}((-\frac 12)^{(\widehat{{\bf c}}_1)}).
\label{fi58}
\end{eqnarray}

Similarly, 
\begin{eqnarray}
\ &&r_{12}^{(1)}=\xi _1^{*}((+\frac 12)^{(\widehat{{\bf d}})};(+\frac 12)^{(%
\widehat{{\bf c}}_1)})\xi _1((-\frac 12)^{(\widehat{{\bf d}})};(+\frac 12)^{(%
\widehat{{\bf c}}_1)})r^{(1)}((+\frac 12)^{(\widehat{{\bf c}}_1)})  \nonumber
\\
\ &&+\xi _1^{*}((+\frac 12)^{(\widehat{{\bf d}})};(-\frac 12)^{(\widehat{%
{\bf c}}_1)})\xi _1((-\frac 12)^{(\widehat{{\bf d}})};(-\frac 12)^{(\widehat{%
{\bf c}}_1)})r^{(1)}((-\frac 12)^{(\widehat{{\bf c}}_1)})  \nonumber \\
\ &=&[-\frac 12\sin \theta _d\cos \theta _1+\frac 12\sin \theta _1\cos
\theta _d\cos (\varphi _d-\varphi _1)  \nonumber \\
&&+\frac i2\sin \theta _1\sin (\varphi _d-\varphi _1)]r^{(1)}((+\frac 12)^{(%
\widehat{{\bf c}}_1)})  \nonumber \\
&&\ +[\frac 12\sin \theta _d\cos \theta _1-\frac 12\sin \theta _1\cos \theta
_d\cos (\varphi _d-\varphi _1)  \nonumber \\
&&\ -\frac i2\sin \theta _1\sin (\varphi _d-\varphi _1)]r^{(1)}((-\frac
12)^{(\widehat{{\bf c}}_1)}),  \label{fi59}
\end{eqnarray}

\begin{equation}
r_{21}^{(1)}=r_{12}^{(1)*}  \label{si60}
\end{equation}
and 
\begin{eqnarray}
&&r_{22}^{(1)}=[\sin ^2(\theta _d-\theta _1)/2+\sin \theta _d\sin \theta
_1\sin ^2(\varphi _d-\varphi _1)/2]r^{(1)}((+\frac 12)^{(\widehat{{\bf c}}%
_1)})  \nonumber \\
&&\ \ \ +[\cos ^2(\theta _d-\theta _1)/2-\sin \theta _d\sin \theta _1\sin
^2(\varphi _d-\varphi _1)/2]r^{(1)}((-\frac 12)^{(\widehat{{\bf c}}_1)}).
\label{si61}
\end{eqnarray}

The elements of $[r^{(2)}]_2$ are identical in form to those of $%
[r^{(1)}]_1. $ The difference is that in system $2$ the vector $\widehat{%
{\bf f}}$ plays the role that the vector $\widehat{{\bf d}}$ plays in system 
$1$. Thus, wherever $d$ appears, it is replaced by $f$. Wherever the label $%
1 $ appears, it is replaced by $2$. Wherever $(\pm \frac 12)^{(\widehat{{\bf %
c}}_1)}$ appears, it is replaced by $(\pm \frac 12)^{(\widehat{{\bf c}}_2)}.$
Finally wherever $r^{(1)}$ appears, it is replaced by $r^{(2)}$. Thus, the
elements of $[r^{(2)}]_2$ are 
\begin{eqnarray}
&&r_{11}^{(2)}=[\cos ^2(\theta _f-\theta _2)/2-\sin \theta _d\sin \theta
_1\sin ^2(\varphi _f-\varphi _2)/2]r^{(2)}((+\frac 12)^{(\widehat{{\bf c}}%
_2)})  \nonumber \\
&&\ \ \ +[\sin ^2(\theta _f-\theta _2)/2+\sin \theta _d\sin \theta _1\sin
^2(\varphi _f-\varphi _2)/2]r^{(2)}((-\frac 12)^{(\widehat{{\bf c}}_2)}),
\label{si62}
\end{eqnarray}

\begin{eqnarray}
\ &&r_{12}^{(2)}=[-\frac 12\sin \theta _f\cos \theta _2+\frac 12\sin \theta
_2\cos \theta _f\cos (\varphi _f-\varphi _2)  \nonumber \\
&&\ +\frac i2\sin \theta _2\sin (\varphi _f-\varphi _2)]r^{(2)}((+\frac
12)^{(\widehat{{\bf c}}_2)})  \nonumber \\
&&\ +[\frac 12\sin \theta _f\cos \theta _2-\frac 12\sin \theta _2\cos \theta
_f\cos (\varphi _f-\varphi _2)  \nonumber \\
&&\ -\frac i2\sin \theta _2\sin (\varphi _f-\varphi _2)]r^{(2)}((-\frac
12)^{(\widehat{{\bf c}}_2)}),  \label{si63}
\end{eqnarray}

\begin{equation}
r_{21}^{(2)}=r_{12}^{(2)*}  \label{si64}
\end{equation}
and

\begin{eqnarray}
&&r_{22}^{(2)}=[\sin ^2(\theta _f-\theta _2)/2+\sin \theta _f\sin \theta
_2\sin ^2(\varphi _f-\varphi _2)/2]r^{(2)}((+\frac 12)^{(\widehat{{\bf c}}%
_2)})  \nonumber \\
&&\ \ \ \ \ +[\cos ^2(\theta _f-\theta _2)/2-\sin \theta _f\sin \theta
_2\sin ^2(\varphi _f-\varphi _2)/2]r^{(2)}((-\frac 12)^{(\widehat{{\bf c}}%
_2)}).  \label{si65}
\end{eqnarray}

\subsection{The Vector States}

\subsubsection{The Triplet State}

We start our calculation of the states by looking at the probability
amplitudes corresponding to the triplet state, defined by the quantum
numbers $s=1,$ $M^{(\widehat{{\bf a}})}=0,\pm 1.$ We first consider the case 
$M^{(\widehat{{\bf a}})}=1.$

\paragraph{The $M^{(\widehat{{\bf a}})}=1$ State}

For this case, we write, 
\begin{equation}
\Psi (1,1^{(\widehat{{\bf a}})};(m_1)_u^{(\widehat{{\bf c}}_1)},(m_2)_v^{(%
\widehat{{\bf c}}_2)})=\Psi (s=1,M=1^{(\widehat{{\bf a}})};(m_1)_u^{(%
\widehat{{\bf c}}_1)},(m_2)_v^{(\widehat{{\bf c}}_2)}),  \label{si66}
\end{equation}
and the generalized probability amplitude is [1] 
\begin{eqnarray}
\ &&\Psi (1,1^{(\widehat{{\bf a}})};(m_1)_u^{(\widehat{{\bf c}}%
_1)},(m_2)_v^{(\widehat{{\bf c}}_2)})=  \nonumber \\
&&\sum_{\alpha ,\alpha ^{\prime }}\left[ \sum_l\zeta (1,1^{(\widehat{{\bf a}}%
)};1,M_l^{(\widehat{{\bf k}})})\vartheta (1,M_l^{(\widehat{{\bf k}}%
)};(m_1)_\alpha ^{(\widehat{{\bf k}})},(m_2)_{\alpha ^{\prime }}^{(\widehat{%
{\bf k}})})\right]  \nonumber \\
\ &&\times \Phi ((m_1)_\alpha ^{(\widehat{{\bf k}})},(m_2)_{\alpha ^{\prime
}}^{(\widehat{{\bf k}})});(m_1)_u^{(\widehat{{\bf c}}_1)},(m_2)_v^{(\widehat{%
{\bf c}}_2)}).  \label{si67}
\end{eqnarray}
Thus, 
\begin{eqnarray}
&&\Psi (1,1^{(\widehat{{\bf a}})};(m_1)_u^{(\widehat{{\bf c}}_1)},(m_2)_v^{(%
\widehat{{\bf c}}_2)})=\sum_{\alpha ,\alpha ^{\prime }}\chi (1,1^{(\widehat{%
{\bf a}})};B_{\alpha \alpha ^{\prime }})  \nonumber \\
&&\times \Phi (B_{\alpha \alpha ^{\prime }};(m_1)_u^{(\widehat{{\bf c}}%
_1)},(m_2)_v^{(\widehat{{\bf c}}_2)}),  \label{si68}
\end{eqnarray}
where

\begin{equation}
\chi (1,1^{(\widehat{{\bf a}})};B_{\alpha \alpha ^{\prime }})=\sum_l\zeta
(1,1^{(\widehat{{\bf a}})};1,M_l^{(\widehat{{\bf k}})})\vartheta (1,M_l^{(%
\widehat{{\bf k}})};B_{\alpha \alpha ^{\prime }}).  \label{si69}
\end{equation}

With the probability amplitude expressed in the form Eq. (\ref{si68}), which
is identical in form to Eq. (\ref{tw27}), the transformation to matrix form
is straightforwardly achieved. We find that the matrix form of the
probability amplitude is 
\begin{equation}
[\Psi (1,1^{(\widehat{{\bf a}})};(m_1)^{(\widehat{{\bf c}}_1)},(m_2)^{(%
\widehat{{\bf c}}_2)})]=\sum_{\alpha ,\alpha ^{\prime }}\chi (1,1^{(\widehat{%
{\bf a}})};B_{\alpha \alpha ^{\prime }})[\eta _1((B_{\alpha \alpha ^{\prime
}})_1)]_1[\eta _2((B_{\alpha \alpha ^{\prime }})_2)]_2.  \label{se70}
\end{equation}

As $B_{\alpha \alpha ^{\prime }}$ takes the values Eqs. (\ref{ni19}) - (\ref
{tw22}), we have

\begin{eqnarray}
&[\Psi (1,1^{(\widehat{{\bf a}})};(m_1)^{(\widehat{{\bf c}}_1)},(m_2)^{(%
\widehat{{\bf c}}_2)})]=&\chi (1,1^{(\widehat{{\bf a}})};(+\tfrac 12)^{(%
\widehat{{\bf k}})},(+\tfrac 12)^{(\widehat{{\bf k}})})  \nonumber \\
&&\times [\eta _1((+\tfrac 12)^{(\widehat{{\bf k}})})]_1[\eta _2((+\tfrac
12)^{(\widehat{{\bf k}})})]_2  \nonumber \\
&&+\chi (1,1^{(\widehat{{\bf a}})};(+\tfrac 12)^{(\widehat{{\bf k}}%
)},(-\tfrac 12)^{(\widehat{{\bf k}})})[\eta _1((+\tfrac 12)^{(\widehat{{\bf k%
}})})]_1[\eta _2((-\tfrac 12)^{(\widehat{{\bf k}})})]_2  \nonumber \\
&&+\chi (1,1^{(\widehat{{\bf a}})};(-\tfrac 12)^{(\widehat{{\bf k}}%
)},(+\tfrac 12)^{(\widehat{{\bf k}})})[\eta _1((-\tfrac 12)^{(\widehat{{\bf k%
}})})]_1[\eta _2((+\tfrac 12)^{(\widehat{{\bf k}})})]_2  \nonumber \\
&&+\chi (1,1^{(\widehat{{\bf a}})};(-\tfrac 12)^{(\widehat{{\bf k}}%
)},(-\tfrac 12)^{(\widehat{{\bf k}})})[\eta _1((-\tfrac 12)^{(\widehat{{\bf k%
}})})]_1[\eta _2((-\tfrac 12)^{(\widehat{{\bf k}})})]_2,  \nonumber \\
&&  \label{se71}
\end{eqnarray}
where

\begin{equation}
\lbrack \eta _1((\pm \tfrac 12)^{(\widehat{{\bf k}})})]_1=\left( 
\begin{array}{c}
\eta _1((\pm \tfrac 12)^{(\widehat{{\bf k}})};(+\tfrac 12)^{(\widehat{{\bf d}%
})}) \\ 
\eta _1((\pm \tfrac 12)^{(\widehat{{\bf k}})};(-\tfrac 12)^{(\widehat{{\bf d}%
})})
\end{array}
\right) _1  \label{se72}
\end{equation}
and 
\begin{equation}
\lbrack \eta _2((\pm \tfrac 12)^{(\widehat{{\bf k}})})]_2=\left( 
\begin{array}{c}
\eta _2((\pm \tfrac 12)^{(\widehat{{\bf k}})});(+\tfrac 12)^{(\widehat{{\bf f%
}})}) \\ 
\eta _2((\pm \tfrac 12)^{(\widehat{{\bf k}})});(-\tfrac 12)^{(\widehat{{\bf f%
}})})
\end{array}
\right) _2.  \label{se73}
\end{equation}

The $\eta _1$'s and $\eta _2$'s are known. They are just the spin-$1/2$
probability amplitudes and are essentially identical to the $\xi $'s, Eqs. (%
\ref{fi53}) - (\ref{fi56}). The only difference is in the direction vectors.
In the labelling of the arguments for the $\eta ^{\prime }$s, the initial
direction corresponds to the $z$ axis, so that its direction vector is $%
\widehat{{\bf k}}$; the final directions are defined by $\widehat{{\bf d}}$
and $\widehat{{\bf f}}$ for system $1$ and system $2$ respectively. From
Eqs. (\ref{fi53}) -(\ref{fi56}), with the arguments appropriately changed,
we get

\begin{equation}
\lbrack \eta _1((+\tfrac 12 )^{(\widehat{{\bf k}})}]_1=\left( 
\begin{array}{c}
\cos \theta _d/2 \\ 
-\sin \theta _d/2
\end{array}
\right) _1\text{,\ \ }  \label{se74}
\end{equation}
\begin{equation}
\text{\ \ }[\eta _1(-\tfrac 12 )^{(\widehat{\QTR{textbf}{k}})}]_1=\left( 
\begin{array}{c}
\sin \theta _d/2e^{-i\varphi _d} \\ 
\cos \theta _d/2e^{-i\varphi _d}
\end{array}
\right) _1,  \label{se74a}
\end{equation}
\begin{equation}
\lbrack \eta _2(+\tfrac 12 )^{(\widehat{{\bf k}})}]_2=\left( 
\begin{array}{c}
\cos \theta _f/2 \\ 
-\sin \theta _f/2
\end{array}
\right) _2  \label{se74b}
\end{equation}
and 
\begin{equation}
\text{\ \ }[\eta _2((-\tfrac 12 )^{(\widehat{\QTR{textbf}{k}})})]_2=\left( 
\begin{array}{c}
\sin \theta _f/2e^{-i\varphi _f} \\ 
\cos \theta _f/2e^{-i\varphi _f}
\end{array}
\right) _2.  \label{se75}
\end{equation}
It only remains to compute the $\chi $'s$.$ According to Eq. (\ref{si69})

\begin{eqnarray}
&&\chi (1,1^{(\widehat{{\bf a}})};(+\tfrac 12)^{(\widehat{{\bf k}}%
)},(+\tfrac 12)^{(\widehat{{\bf k}})})=\chi (1,1^{(\widehat{{\bf a}}%
)};(+\tfrac 12)^{(\widehat{{\bf k}})},(+\tfrac 12)^{(\widehat{{\bf k}})}) 
\nonumber \\
&=&\zeta (1,1^{(\widehat{{\bf a}})};1,1^{(\widehat{{\bf k}})})\vartheta
(1,1^{(\widehat{{\bf k}})};(+\tfrac 12)^{(\widehat{{\bf k}})},(+\tfrac 12)^{(%
\widehat{{\bf k}})})  \nonumber \\
&&+\zeta (1,1^{(\widehat{{\bf a}})};1,0^{(\widehat{{\bf k}})})\vartheta
(1,0^{(\widehat{{\bf k}})};(+\tfrac 12)^{(\widehat{{\bf k}})},(+\tfrac 12)^{(%
\widehat{{\bf k}})})  \nonumber \\
&&+\zeta (1,1^{(\widehat{{\bf a}})};1,(-1)^{(\widehat{{\bf k}})})\vartheta
(1,(-1)^{(\widehat{{\bf k}})};(+\tfrac 12)^{(\widehat{{\bf k}})},(+\tfrac
12)^{(\widehat{{\bf k}})}).  \label{se76}
\end{eqnarray}

The angles defining $\widehat{{\bf a}}$ are $(\theta ,\varphi ).$ As we have
shown [1,9],

\begin{equation}
\zeta (1,1^{(\widehat{{\bf a}})};1,1^{(\widehat{{\bf k}})})=\cos ^2\theta
/2e^{-i\varphi }  \label{se77}
\end{equation}

\begin{equation}
\zeta (1,1^{(\widehat{{\bf a}})};1,0^{(\widehat{{\bf k}})})=\frac 1{\sqrt{2}%
}\sin \theta  \label{se78}
\end{equation}
and

\begin{equation}
\zeta (1,1^{(\widehat{{\bf a}})};1,(-1)^{(\widehat{{\bf k}})})=\sin ^2\theta
/2e^{i\varphi }.  \label{se79}
\end{equation}

The $\vartheta $'s are Clebsch-Gordan coefficients. As a result 
\begin{equation}
\vartheta (1,1^{(\widehat{{\bf k}})};(+\tfrac 12)^{(\widehat{{\bf k}}%
)},(+\tfrac 12)^{(\widehat{{\bf k}})})=C(\tfrac 12\tfrac 121;\tfrac 12\tfrac
121)=1,  \label{ei80}
\end{equation}

\begin{equation}
\vartheta (1,0^{(\widehat{{\bf k}})};(+\tfrac 12)^{(\widehat{{\bf k}}%
)},(+\tfrac 12)^{(\widehat{{\bf k}})})=C(\tfrac 12\tfrac 121;\tfrac 12\tfrac
120)=0  \label{ei81}
\end{equation}
and 
\begin{equation}
\vartheta (1,(-1)^{(\widehat{{\bf k}})};(+\tfrac 12)^{(\widehat{{\bf k}}%
)},(+\tfrac 12)^{(\widehat{{\bf k}})})=C(\tfrac 12\tfrac 121;\tfrac 12\tfrac
12-1)=0.  \label{ei82}
\end{equation}

Hence,

\begin{equation}
\chi (1,1^{(\widehat{{\bf a}})};(+\tfrac 12)^{(\widehat{{\bf k}})},(+\tfrac
12)^{(\widehat{{\bf k}})})=\frac 12\sin \theta .  \label{ei83}
\end{equation}

Similarly, 
\begin{eqnarray}
&&\chi (1,1^{(\widehat{{\bf a}})};(+\tfrac 12)^{(\widehat{{\bf k}}%
)},(-\tfrac 12)^{(\widehat{{\bf k}})})=\zeta (1,1^{(\widehat{{\bf a}}%
)};1,1^{(\widehat{{\bf k}})})\vartheta (1,1^{(\widehat{{\bf k}})};(+\tfrac
12)^{(\widehat{{\bf k}})},(-\tfrac 12)^{(\widehat{{\bf k}})})  \nonumber \\
&&+\zeta (1,1^{(\widehat{{\bf a}})};1,0^{(\widehat{{\bf k}})})\vartheta
(1,0^{(\widehat{{\bf k}})};(+\tfrac 12)^{(\widehat{{\bf k}})},(-\tfrac 12)^{(%
\widehat{{\bf k}})})  \nonumber \\
&&+\zeta (1,1^{(\widehat{{\bf a}})};1,(-1)^{(\widehat{{\bf k}})})\vartheta
(1,(-1)^{(\widehat{{\bf k}})};(+\tfrac 12)^{(\widehat{{\bf k}})},(-\tfrac
12)^{(\widehat{{\bf k}})}).  \label{ei84}
\end{eqnarray}

This is the same as the expression for $\chi (1,1^{(\widehat{{\bf a}}%
)};(+\tfrac 12)^{(\widehat{{\bf k}})},(+\tfrac 12)^{(\widehat{{\bf k}})})$,
except for the change in the $\vartheta $'s. Since 
\begin{equation}
\vartheta (1,+1^{(\widehat{{\bf k}})};(+\tfrac 12)^{(\widehat{{\bf k}}%
)},(-\tfrac 12)^{(\widehat{{\bf k}})})=C(\tfrac 12\tfrac 121;\tfrac
12,-\tfrac 121)=0,  \label{ei85}
\end{equation}
\begin{equation}
\vartheta (1,(-1)^{(\widehat{{\bf k}})};(+\tfrac 12)^{(\widehat{{\bf k}}%
)},(-\tfrac 12)^{(\widehat{{\bf k}})})=C(\tfrac 12\tfrac 121;\tfrac
12,-\tfrac 12,-1)=0  \label{ei86}
\end{equation}
and 
\begin{equation}
\vartheta (1,0^{(\widehat{{\bf k}})};(+\tfrac 12)^{(\widehat{{\bf k}}%
)},(-\tfrac 12)^{(\widehat{{\bf k}})})=C(\tfrac 12\tfrac 121;\tfrac
12,-\tfrac 120)=\frac 1{\sqrt{2}},  \label{ei87}
\end{equation}
we find that 
\begin{equation}
\chi (1,1^{(\widehat{{\bf a}})};(+\tfrac 12)^{(\widehat{{\bf k}})},(-\tfrac
12)^{(\widehat{{\bf k}})})=\frac 12\sin \theta .  \label{ei88}
\end{equation}

In the same way 
\begin{eqnarray}
&&\chi (1,+1^{(\widehat{{\bf a}})};(-\tfrac 12)^{(\widehat{{\bf k}}%
)},(+\tfrac 12)^{(\widehat{{\bf k}})})=\zeta (1,+1^{(\widehat{{\bf a}}%
)};1,+1^{(\widehat{{\bf k}})})\vartheta (1,+1^{(\widehat{{\bf k}})};(-\tfrac
12)^{(\widehat{{\bf k}})},(+\tfrac 12)^{(\widehat{{\bf k}})})  \nonumber \\
&&+\zeta (1,+1^{(\widehat{{\bf a}})};1,0^{(\widehat{{\bf k}})})\vartheta
(1,0^{(\widehat{{\bf k}})};(-\tfrac 12)^{(\widehat{{\bf k}})},(+\tfrac 12)^{(%
\widehat{{\bf k}})})  \nonumber \\
&&+\zeta (1,+1^{(\widehat{{\bf a}})};1,(-1)^{(\widehat{{\bf k}}%
)}{})\vartheta (1,(-1)^{(\widehat{{\bf k}})};(-\tfrac 12)^{(\widehat{{\bf k}}%
)},(+\tfrac 12)^{(\widehat{{\bf k}})}).  \label{ei89}
\end{eqnarray}

In this case, we have

\begin{equation}
\vartheta (1,+1^{(\widehat{{\bf a}})};(-\tfrac 12)^{(\widehat{{\bf k}}%
)},(+\tfrac 12)^{(\widehat{{\bf k}})})=C(\tfrac 12\tfrac 121;-\tfrac
12\tfrac 121)=0,  \label{ni90}
\end{equation}
\begin{equation}
\vartheta (1,(-1)^{(\widehat{{\bf a}})};(-\tfrac 12)^{(\widehat{{\bf k}}%
)},(+\tfrac 12)^{(\widehat{{\bf k}})})=C(\tfrac 12\tfrac 121;-\tfrac
12\tfrac 12,-1)=0  \label{ni91}
\end{equation}
and 
\begin{equation}
\vartheta (1,0^{(\widehat{{\bf a}})};(-\tfrac 12)^{(\widehat{{\bf k}}%
)},(+\tfrac 12)^{(\widehat{{\bf k}})})=C(\tfrac 12\tfrac 121;-\tfrac
12\tfrac 120)=\frac 1{\sqrt{2}}.  \label{ni92}
\end{equation}

Thus, 
\begin{equation}
\chi (1,+1^{(\widehat{{\bf a}})};(-\tfrac 12)^{(\widehat{{\bf k}})},(+\tfrac
12)^{(\widehat{{\bf k}})})=\frac 12\sin \theta .  \label{ni93}
\end{equation}

Finally 
\begin{eqnarray}
&&\chi (1,1^{(\widehat{{\bf a}})}{};{}(-\tfrac 12)^{(\widehat{{\bf k}}%
)},(-\tfrac 12)^{(\widehat{{\bf k}})})=  \nonumber \\
&&\zeta (1,1^{(\widehat{{\bf a}})};1,1^{(\widehat{{\bf k}})})\vartheta
(1,+1^{(\widehat{{\bf k}})};(-\tfrac 12)^{(\widehat{{\bf k}})},(-\tfrac
12)^{(\widehat{{\bf k}})})  \nonumber \\
&&+\zeta (1,1^{(\widehat{{\bf a}})};1,0^{(\widehat{{\bf k}})})\vartheta
(1,0^{(\widehat{{\bf k}})};(-\tfrac 12)^{(\widehat{{\bf k}})},(-\tfrac 12)^{(%
\widehat{{\bf k}})})  \nonumber \\
&&+\zeta (1,1^{(\widehat{{\bf a}})};1,(-1)^{(\widehat{{\bf k}})})\vartheta
(1,(-1)^{(\widehat{{\bf k}})};(-\tfrac 12)^{(\widehat{{\bf k}})},(-\tfrac
12)^{(\widehat{{\bf k}})}).  \label{ni94}
\end{eqnarray}

With

\begin{equation}
\vartheta (1,1^{(\widehat{{\bf k}})};(-\tfrac 12)^{(\widehat{{\bf k}}%
)},(-\tfrac 12)^{(\widehat{{\bf k}})})=C(\tfrac 12\tfrac 121;-\tfrac
12,-\tfrac 121)=0,  \label{ni95}
\end{equation}
\begin{equation}
\vartheta (1,(-1)^{(\widehat{{\bf k}})};(-\tfrac 12)^{(\widehat{{\bf k}}%
)},(-\tfrac 12)^{(\widehat{{\bf k}})})=C(\tfrac 12\tfrac 121;-\tfrac
12,-\tfrac 12,-1)=1  \label{ni96}
\end{equation}
and 
\begin{equation}
\vartheta (1,0^{(\widehat{{\bf k}})};(-\tfrac 12)^{(\widehat{{\bf k}}%
)},(-\tfrac 12)^{(\widehat{{\bf k}})})=C(\tfrac 12\tfrac 121;-\tfrac
12,-\tfrac 120)=0.  \label{ni97}
\end{equation}
we obtain 
\begin{equation}
\chi (1,1^{(\widehat{{\bf a}})};(-\tfrac 12)^{(\widehat{{\bf k}})},(-\tfrac
12)^{(\widehat{{\bf k}})})=\sin ^2\theta /2e^{i\varphi }.  \label{ni98}
\end{equation}

Combining all these results together, we find that the matrix state for $s=1$%
, $M^{(\widehat{{\bf a}})}=1$ is 
\begin{eqnarray}
&[\Psi (1,1^{(\widehat{{\bf a}})};(m_1)^{\widehat{{\bf c}}_1},(m_2)^{%
\widehat{{\bf c}}_2})]=&\cos ^2\theta /2e^{-i\varphi }[\eta _1((+\tfrac
12)^{(\widehat{{\bf k}})})]_1[\eta _2((+\tfrac 12)^{(\widehat{{\bf k}})})]_2
\nonumber \\
&&+\frac 12\sin \theta [\eta _1((+\tfrac 12)^{(\widehat{{\bf k}})})]_1[\eta
_2((-\tfrac 12)^{(\widehat{{\bf k}})})]_2  \nonumber \\
&&+\frac 12\sin \theta [\eta _1((-\tfrac 12)^{(\widehat{{\bf k}})})]_1[\eta
_2((+\tfrac 12)^{(\widehat{{\bf k}})})]_2  \nonumber \\
&&+\sin ^2\theta /2e^{i\varphi }[\eta _1((-\tfrac 12)^{(\widehat{{\bf k}}%
)})]_1[\eta _2((-\tfrac 12)^{(\widehat{{\bf k}})})]_2.  \nonumber \\
&&  \label{ni99}
\end{eqnarray}

Thus the generalised form of the triplet state for $M^{(\widehat{{\bf a}}%
)}=1 $ is, 
\begin{eqnarray}
&[\Psi (1,1^{(\widehat{{\bf a}})};(m_1)^{(\widehat{{\bf c}}_1)},(m_2)^{(%
\widehat{{\bf c}}_2)})]=&\cos ^2\theta /2e^{-i\varphi }\left( 
\begin{array}{c}
\cos \theta _d/2 \\ 
-\sin \theta _d/2
\end{array}
\right) _1\left( 
\begin{array}{c}
\cos \theta _f/2 \\ 
-\sin \theta _f/2
\end{array}
\right) _2  \nonumber \\
\ \ \ &&+\frac 12\sin \theta \left( 
\begin{array}{c}
\cos \theta _d/2 \\ 
-\sin \theta _d/2
\end{array}
\right) _1\left( 
\begin{array}{c}
\sin \theta _f/2e^{-i\varphi _f} \\ 
\cos \theta _f/2e^{-i\varphi _f}
\end{array}
\right) _2  \nonumber \\
&&\ \ +\frac 12\sin \theta \left( 
\begin{array}{c}
\sin \theta _d/2e^{-i\varphi _d} \\ 
\cos \theta _d/2e^{-i\varphi _d}
\end{array}
\right) _1\left( 
\begin{array}{c}
\cos \theta _f/2 \\ 
-\sin \theta _f/2
\end{array}
\right) _2  \nonumber \\
\ \ \ &&+\sin ^2\theta /2e^{i\varphi }\left( 
\begin{array}{c}
\sin \theta _d/2e^{-i\varphi _d} \\ 
\cos \theta _d/2e^{-i\varphi _d}
\end{array}
\right) _1\left( 
\begin{array}{c}
\sin \theta _f/2e^{-i\varphi _f} \\ 
\cos \theta _f/2e^{-i\varphi _f}
\end{array}
\right) _2.  \nonumber \\
&&  \label{hu100}
\end{eqnarray}

\paragraph{The $M^{(\widehat{{\bf a}})}=0$ State}

For this case, the probability amplitude is [1]

\begin{eqnarray}
\ &&\Psi (1,0^{(\widehat{{\bf a}})};(m_1)_u^{(\widehat{{\bf c}}%
_1)},(m_2)_v^{(\widehat{{\bf c}}_2)})=\sum_{\alpha ,\alpha ^{\prime
}}\{\sum_l\zeta (1,0^{(\widehat{{\bf a}})};1,M_l^{(\widehat{{\bf k}})}) 
\nonumber \\
\ &&\times \vartheta (1,M_l^{(\widehat{{\bf k}})};(m_1)_\alpha ^{(\widehat{%
{\bf k}})},(m_2)_{\alpha ^{\prime }}^{(\widehat{{\bf k}})})\}  \nonumber \\
\ &&\times \Phi ((m_1)_\alpha ^{(\widehat{{\bf k}})},(m_2)_{\alpha ^{\prime
}}^{(\widehat{{\bf k}})};(m_1)_u^{(\widehat{{\bf c}}_1)},(m_2)_v^{(\widehat{%
{\bf c}}_2)})  \nonumber \\
\ &=&\sum_j\chi (1,0^{(\widehat{{\bf a}})};B_{\alpha \alpha ^{\prime }})\Phi
(B_{\alpha \alpha ^{\prime }};(m_1)_u^{(\widehat{{\bf c}}_1)},(m_2)_v^{(%
\widehat{{\bf c}}_2)}),  \label{hu101}
\end{eqnarray}
where 
\begin{equation}
\chi (1,0^{(\widehat{{\bf a}})};B_{\alpha \alpha ^{\prime }})=\sum_l\zeta
(1,0^{(\widehat{{\bf a}})};1,M_l^{(\widehat{{\bf k}})})\vartheta (1,M_l^{(%
\widehat{{\bf k}})};B_{\alpha \alpha ^{\prime }}).  \label{hu102}
\end{equation}
The $\zeta $'s change because they are functions of the initial-state
quantum numbers. Thus, we have [9]

\begin{equation}
\zeta (1,0^{(\widehat{{\bf a}})};1,1^{(\widehat{{\bf k}})})=-\frac 1{\sqrt{2}%
}\sin \theta e^{-i\varphi }  \label{hu103}
\end{equation}

\begin{equation}
\zeta (1,0^{(\widehat{{\bf a}})};1,0^{(\widehat{{\bf k}})})=\cos \theta
\label{hu104}
\end{equation}
and

\begin{equation}
\zeta (1,0^{(\widehat{{\bf a}})};1,(-1)^{(\widehat{{\bf k}})})=\frac 1{\sqrt{%
2}}\sin \theta e^{i\varphi }.  \label{hu105}
\end{equation}
However, the $\vartheta $'s do not change. Thus, using Eqs. (\ref{ei80}) - (%
\ref{ei82}) for the $\vartheta $'s, we find that 
\begin{eqnarray}
&&\chi (1,0^{(\widehat{{\bf a}})};(+\tfrac 12)^{(\widehat{{\bf k}}%
)},(+\tfrac 12)^{(\widehat{{\bf k}})})=  \nonumber \\
&&\zeta (1,0^{(\widehat{{\bf a}})};1,1^{(\widehat{{\bf k}})})\vartheta
(1,1^{(\widehat{{\bf k}})};(+\tfrac 12)^{(\widehat{{\bf k}})},(+\tfrac 12)^{(%
\widehat{{\bf k}})})  \nonumber \\
&&+\zeta (1,0^{(\widehat{{\bf a}})};1,0^{(\widehat{{\bf k}})})\vartheta
(1,0^{(\widehat{{\bf k}})};(+\tfrac 12)^{(\widehat{{\bf k}})},(+\tfrac 12)^{(%
\widehat{{\bf k}})})  \nonumber \\
&&+\zeta (1,0^{(\widehat{{\bf a}})};1,(-1)^{(\widehat{{\bf k}})})\vartheta
(1,(-1)^{(\widehat{{\bf k}})};(+\tfrac 12)^{(\widehat{{\bf k}})},(+\tfrac
12)^{(\widehat{{\bf k}})})  \nonumber \\
&=&-\frac 1{\sqrt{2}}\sin \theta e^{-i\varphi }.  \label{hu106}
\end{eqnarray}
The other $\chi $'s are found to be

\begin{eqnarray}
\chi (1,0^{(\widehat{{\bf a}})};(+\tfrac 12)^{(\widehat{{\bf k}})},(-\tfrac
12)^{(\widehat{{\bf k}})}) &=&\chi (1,0^{(\widehat{{\bf a}})};(-\tfrac 12)^{(%
\widehat{{\bf k}})},(+\tfrac 12)^{(\widehat{{\bf k}})})  \nonumber \\
&=&\frac 1{\sqrt{2}}\cos \theta  \label{hu108}
\end{eqnarray}
and 
\begin{equation}
\chi (1,0^{(\widehat{{\bf a}})};(-\tfrac 12)^{(\widehat{{\bf k}})},(-\tfrac
12)^{(\widehat{{\bf k}})})=\frac 1{\sqrt{2}}\sin \theta e^{i\varphi }.
\label{hu109}
\end{equation}

Thus 
\begin{eqnarray}
&[\Psi (1,0^{(\widehat{{\bf a}})};(m_1)^{(\widehat{{\bf c}}_1)},(m_2)^{(%
\widehat{{\bf c}}_2)}]=&-\frac 1{\sqrt{2}}\sin \theta e^{-i\varphi }\left( 
\begin{array}{c}
\cos \theta _d/2 \\ 
-\sin \theta _d/2
\end{array}
\right) _1\left( 
\begin{array}{c}
\cos \theta _f/2 \\ 
-\sin \theta _f/2
\end{array}
\right) _2  \nonumber \\
\ \ \ \ \ &&+\frac 1{\sqrt{2}}\cos \theta \left( 
\begin{array}{c}
\cos \theta _d/2 \\ 
-\sin \theta _d/2
\end{array}
\right) _1\left( 
\begin{array}{c}
\sin \theta _f/2e^{-i\varphi _f} \\ 
\cos \theta _f/2e^{-i\varphi _f}
\end{array}
\right) _2  \nonumber \\
&&\ +\frac 1{\sqrt{2}}\cos \theta \left( 
\begin{array}{c}
\sin \theta _d/2e^{-i\varphi _d} \\ 
\cos \theta _d/2e^{-i\varphi _d}
\end{array}
\right) _1\left( 
\begin{array}{c}
\cos \theta _f/2 \\ 
-\sin \theta _f/2
\end{array}
\right) _2  \nonumber \\
\ \ \ \ \ &&+\frac 1{\sqrt{2}}\sin \theta e^{i\varphi }\left( 
\begin{array}{c}
\sin \theta _d/2e^{-i\varphi _d} \\ 
\cos \theta _d/2e^{-i\varphi _d}
\end{array}
\right) _1\left( 
\begin{array}{c}
\sin \theta _f/2e^{-i\varphi _f} \\ 
\cos \theta _f/2e^{-i\varphi _f}
\end{array}
\right) _2.  \nonumber \\
&&  \label{hu110}
\end{eqnarray}

\paragraph{The $M^{(\widehat{{\bf a}})}=-1$ State}

For this case, the probability amplitude is 
\begin{eqnarray}
\ &&\Psi (1,(-1)^{(\widehat{{\bf a}})};(m_1)_u^{(\widehat{{\bf c}}%
_1)},(m_2)_v^{(\widehat{{\bf c}}_2)})=\sum_{\alpha ,\alpha ^{\prime
}}\{\sum_l\zeta (1,(-1)^{(\widehat{{\bf a}})};1,M_l^{(\widehat{{\bf k}})}) 
\nonumber \\
\ &&\times \vartheta (1,M_l^{(\widehat{{\bf k}})};(m_1)_\alpha ^{(\widehat{%
{\bf k}})},(m_2)_{\alpha ^{\prime }}^{(\widehat{{\bf k}})})\}  \nonumber \\
\ &&\times \Phi ((m_1)_\alpha ^{(\widehat{{\bf k}})},(m_2)_{\alpha ^{\prime
}}^{(\widehat{{\bf k}})};(m_1)_u^{(\widehat{{\bf c}}_1)},(m_2)_v^{(\widehat{%
{\bf c}}_2)})  \nonumber \\
\ &=&\sum_{\alpha ,\alpha ^{\prime }}\chi (1,(-1)^{(\widehat{{\bf a}}%
)};B_{\alpha \alpha ^{\prime }})\Phi (B_{\alpha \alpha ^{\prime }};(m_1)_u^{(%
\widehat{{\bf c}}_1)},(m_2)_v^{(\widehat{{\bf c}}_2)}),  \label{hu111}
\end{eqnarray}
where

\begin{equation}
\ \chi (1,(-1)^{(\widehat{{\bf a}})};B_{\alpha \alpha ^{\prime }})=\chi
(1,(-1)^{(\widehat{{\bf a}})};B_{\alpha \alpha ^{\prime }}):=\sum_l\zeta
(1,(-1)^{(\widehat{{\bf a}})};1,M_l^{(\widehat{{\bf k}})})\vartheta (1,M_l^{(%
\widehat{{\bf k}})};B_{\alpha \alpha ^{\prime }})  \label{hu112}
\end{equation}

The $\zeta $'s for this case are

\begin{equation}
\zeta (1,(-1)^{(\widehat{{\bf a}})};1,1^{(\widehat{{\bf k}})})=-\sin
^2\theta /2e^{-i\varphi },  \label{hu113}
\end{equation}

\begin{equation}
\zeta (1,(-1)^{(\widehat{{\bf a}})};1,0^{(\widehat{{\bf k}})})=\frac 1{\sqrt{%
2}}\sin \theta  \label{hu114}
\end{equation}
and

\begin{equation}
\zeta (1,(-1)^{(\widehat{{\bf a}})};1,(-1)^{(\widehat{{\bf k}})})=-\cos
^2\tfrac \theta 2e^{i\varphi }.  \label{hu115}
\end{equation}
Since the $\vartheta $'s remain the same, it follows that 
\begin{equation}
\chi (1,(-1)^{(\widehat{{\bf a}})};(+\tfrac 12)^{(\widehat{{\bf k}}%
)},(+\tfrac 12)^{(\widehat{{\bf k}})}))=-\sin ^2\theta /2e^{-i\varphi },
\label{hu116}
\end{equation}

\begin{eqnarray}
&&\chi (1,(-1)^{(\widehat{{\bf a}})};(+\tfrac 12)^{(\widehat{{\bf k}}%
)},(-\tfrac 12)^{(\widehat{{\bf k}})}))=\chi (1,(-1)^{(\widehat{{\bf a}}%
)};(-\tfrac 12)^{(\widehat{{\bf k}})},(+\tfrac 12)^{(\widehat{{\bf k}})})) 
\nonumber \\
&=&\frac 12\sin \theta ,  \label{hu117}
\end{eqnarray}
and

\begin{equation}
\chi (1,(-1)^{(\widehat{{\bf a}})};(-\tfrac 12)^{(\widehat{{\bf k}}%
)},(-\tfrac 12)^{(\widehat{{\bf k}})}))=-\cos ^2\theta /2e^{i\varphi }.
\label{hu118}
\end{equation}
As a result, we obtain for the matrix state 
\begin{eqnarray}
&[\Psi (1,(-1)^{(\widehat{{\bf a}})};(m_1)^{(\widehat{{\bf c}}_1)},(m_2)^{(%
\widehat{{\bf c}}_2)})]=&-\sin ^2\theta /2e^{-i\varphi }\left( 
\begin{array}{c}
\cos \theta _d/2 \\ 
-\sin \theta _d/2
\end{array}
\right) _1\left( 
\begin{array}{c}
\cos \theta _f/2 \\ 
-\sin \theta _f/2
\end{array}
\right) _2  \nonumber \\
\ \ \ \ \ \ &&\ +\frac 12\sin \theta \left( 
\begin{array}{c}
\cos \theta _d/2 \\ 
-\sin \theta _d/2
\end{array}
\right) _1\left( 
\begin{array}{c}
\sin \theta _f/2e^{-i\varphi _f} \\ 
\cos \theta _f/2e^{-i\varphi _f}
\end{array}
\right) _2  \nonumber \\
&&\ \ +\frac 12\sin \theta \left( 
\begin{array}{c}
\sin \theta _d/2e^{-i\varphi _d} \\ 
\cos \theta _d/2e^{-i\varphi _d}
\end{array}
\right) _1\left( 
\begin{array}{c}
\cos \theta _f/2 \\ 
-\sin \theta _f/2
\end{array}
\right) _2  \nonumber \\
\ \ \ \ \ \ \ &&-\cos ^2\theta /2e^{i\varphi }\left( 
\begin{array}{c}
\sin \theta _d/2e^{-i\varphi _d} \\ 
\cos \theta _d/2e^{-i\varphi _d}
\end{array}
\right) _1\left( 
\begin{array}{c}
\sin \theta _f/2e^{-i\varphi _f} \\ 
\cos \theta _f/2e^{-i\varphi _f}
\end{array}
\right) _2.  \nonumber \\
&&  \label{hu119}
\end{eqnarray}

\subsubsection{The Singlet State}

Having obtained the triplet states, we now seek the singlet state. The
general formula is Eq. (\ref{fi51}). The generalized probability amplitude
for the singlet state is

\begin{eqnarray}
\ &&\Psi (s=0,M=0;(m_1)_u^{(\widehat{{\bf c}}_1)},(m_2)_v^{(\widehat{{\bf c}}%
_2)})=\Psi (0,0;(m_1)_u^{(\widehat{{\bf c}}_1)},(m_2)_v^{(\widehat{{\bf c}}%
_2)})  \nonumber \\
\ &=&\sum_{\alpha ,\alpha ^{\prime }}\chi (0,0;(m_1)_\alpha ^{(\widehat{{\bf %
k}})},(m_2)_{\alpha ^{\prime }}^{(\widehat{{\bf k}})})\Phi ((m_1)_\alpha ^{(%
\widehat{{\bf k}})},(m_2)_{\alpha ^{\prime }}^{(\widehat{{\bf k}}%
)};(m_1)_u^{(\widehat{{\bf c}}_1)},(m_2)_v^{(\widehat{{\bf c}}_2)}).
\label{hu120}
\end{eqnarray}

The $\chi $'s are now directly Clebsch-Gordan coefficients for the case of
total spin $0$ and subspins $1/2$ and $1/2$. Thus,

\begin{equation}
\chi (0,0;(+\tfrac 12)^{(\widehat{{\bf k}})},(+\tfrac 12)^{(\widehat{{\bf k}}%
)})=C(\tfrac 12\tfrac 120,\tfrac 12\tfrac 120)=0,  \label{hu121}
\end{equation}
\begin{equation}
\chi (0,0;(+\tfrac 12)^{(\widehat{{\bf k}})},(-\tfrac 12)^{(\widehat{{\bf k}}%
)})=C(\tfrac 12\tfrac 120,\tfrac 12,-\tfrac 120)=\dfrac 1{\sqrt{2}},
\label{hu122}
\end{equation}
\begin{equation}
\chi (0,0;(-\tfrac 12)^{(\widehat{{\bf k}})},(+\tfrac 12)^{(\widehat{{\bf k}}%
)})=C(\tfrac 12\tfrac 120,-\tfrac 12\tfrac 120)=-\dfrac 1{\sqrt{2}}
\label{hu123}
\end{equation}
and 
\begin{equation}
\chi (0,0;(-\tfrac 12)^{(\widehat{{\bf k}})},(-\tfrac 12)^{(\widehat{{\bf k}}%
)})=C(\tfrac 12\tfrac 120,-\tfrac 12,-\tfrac 120)=0.  \label{hu124}
\end{equation}
This means that the generalized probability amplitude is 
\begin{eqnarray}
&&\Psi (0,0;(m_1)_u^{(\widehat{{\bf c}}_1)},(m_2)_v^{(\widehat{{\bf c}}%
_2)})=\tfrac 1{\sqrt{2}}\Phi ((+\tfrac 12)^{(\widehat{{\bf k}})},(-\tfrac
12)^{(\widehat{{\bf k}})};(m_1)_u^{(\widehat{{\bf c}}_1)},(m_2)_v^{(\widehat{%
{\bf c}}_2)})  \nonumber \\
&&-\tfrac 1{\sqrt{2}}\Phi ((-\tfrac 12)^{(\widehat{{\bf k}})},(+\tfrac 12)^{(%
\widehat{{\bf k}})};(m_1)_u^{(\widehat{{\bf c}}_1)},(m_2)_v^{(\widehat{{\bf c%
}}_2)}).  \label{hu125}
\end{eqnarray}

Hence, the matrix state is

\begin{eqnarray}
&[\Psi (0,0;(m_1)^{(\widehat{{\bf c}}_1)},(m_2)^{(\widehat{{\bf c}}%
_2)})]=&\tfrac 1{\sqrt{2}}[\eta _1((+\tfrac 12)^{(\widehat{{\bf k}}%
)})]_1[\eta _2((-\tfrac 12)^{(\widehat{{\bf k}})})]_2  \nonumber \\
&&-\tfrac 1{\sqrt{2}}[\eta _1((-\tfrac 12)^{(\widehat{{\bf k}})})]_1[\eta
_2((+\tfrac 12)^{(\widehat{{\bf k}})})]_2.  \nonumber \\
&&  \label{hu126}
\end{eqnarray}

Therefore, the generalized matrix form for the singlet state is 
\begin{eqnarray}
&[\Psi (0,0;(m_1)^{(\widehat{{\bf c}}_1)},(m_2)^{(\widehat{{\bf c}}%
_2)})]=&\frac 1{\sqrt{2}}\left( 
\begin{array}{c}
\cos \theta _d/2 \\ 
-\sin \theta _d/2
\end{array}
\right) _1\left( 
\begin{array}{c}
\sin \theta _f/2e^{-i\varphi _f} \\ 
\cos \theta _f/2e^{-i\varphi _f}
\end{array}
\right) _2  \nonumber \\
&&\ \ -\frac 1{\sqrt{2}}\left( 
\begin{array}{c}
\sin \theta _d/2e^{-i\varphi _d} \\ 
\cos \theta _d/2e^{-i\varphi _d}
\end{array}
\right) _1\left( 
\begin{array}{c}
\cos \theta _f/2 \\ 
-\sin \theta _f/2
\end{array}
\right) _2.  \nonumber \\
&&  \label{hu129}
\end{eqnarray}

\section{Recovery of Standard Results}

It is now easy to see how the standard results come about from the current
ones. First of all, if $\widehat{{\bf a}}$ is along the $z$ axis, so that $%
\theta =\varphi =0$, the triplet states become. 
\begin{equation}
\lbrack \Psi (1,1^{(\widehat{{\bf a}})};(m_1)^{(\widehat{{\bf c}}%
_1)},(m_2)^{(\widehat{{\bf c}}_2)}]=\left( 
\begin{array}{c}
\cos \theta _d/2 \\ 
-\sin \theta _d/2
\end{array}
\right) _1\left( 
\begin{array}{c}
\cos \theta _f/2 \\ 
-\sin \theta _f/2
\end{array}
\right) _2,  \label{hu130}
\end{equation}

\begin{eqnarray}
&[\Psi (1,0^{(\widehat{{\bf a}})};(m_1)^{(\widehat{{\bf c}}_1)},(m_2)^{(%
\widehat{{\bf c}}_2)})]=&\frac 1{\sqrt{2}}\ \left( 
\begin{array}{c}
\cos \theta _d/2 \\ 
-\sin \theta _d/2
\end{array}
\right) _1\left( 
\begin{array}{c}
\sin \theta _f/2e^{-i\varphi _f} \\ 
\cos \theta _f/2e^{-i\varphi _f}
\end{array}
\right) _2  \nonumber \\
&&\ \ +\frac 1{\sqrt{2}}\left( 
\begin{array}{c}
\sin \theta _d/2e^{-i\varphi _d} \\ 
\cos \theta _d/2e^{-i\varphi _d}
\end{array}
\right) _1\left( 
\begin{array}{c}
\cos \theta _f/2 \\ 
-\sin \theta _f/2
\end{array}
\right) _2  \nonumber \\
&&  \label{hu131}
\end{eqnarray}
and 
\begin{equation}
\lbrack \Psi (1,(-1)^{(\widehat{{\bf a}})};(m_1)^{(\widehat{{\bf c}}%
_1)},(m_2)^{(\widehat{{\bf c}}_2)})]=-\left( 
\begin{array}{c}
\sin \theta _d/2e^{-i\varphi _d} \\ 
\cos \theta _d/2e^{-i\varphi _d}
\end{array}
\right) _1\left( 
\begin{array}{c}
\sin \theta _f/2e^{-i\varphi _f} \\ 
\cos \theta _f/2e^{-i\varphi _f}
\end{array}
\right) _2,  \label{hu132}
\end{equation}
while the singlet state remains

\begin{eqnarray}
&[\Psi (0,0;(m_1)^{(\widehat{{\bf c}}_1)},(m_2)^{(\widehat{{\bf c}}%
_2)})]=&\dfrac 1{\sqrt{2}}\left( 
\begin{array}{c}
\cos \theta _d/2 \\ 
-\sin \theta _d/2
\end{array}
\right) _1\left( 
\begin{array}{c}
\sin \theta _f/2e^{-i\varphi _f} \\ 
\cos \theta _f/2e^{-i\varphi _f}
\end{array}
\right) _2  \nonumber \\
&&\ \ -\dfrac 1{\sqrt{2}}\left( 
\begin{array}{c}
\sin \theta _d/2e^{-i\varphi _d} \\ 
\cos \theta _d/2e^{-i\varphi _d}
\end{array}
\right) _1\left( 
\begin{array}{c}
\cos \theta _f/2 \\ 
-\sin \theta _f/2
\end{array}
\right) _2.  \nonumber \\
&&  \label{hu133}
\end{eqnarray}
The operator, Eqs. (\ref{fi57}) - (\ref{si65}), remains unchanged.

We recover the standard formulas if in addition, $\widehat{{\bf d}}=\widehat{%
{\bf f}}=\widehat{{\bf k}}$ : 
\begin{equation}
\lbrack \Psi (1,1^{(\widehat{{\bf a}})};(m_1)^{(\widehat{{\bf c}}%
_1)},(m_2)^{(\widehat{{\bf c}}_2)})]=\left( 
\begin{array}{c}
1 \\ 
0
\end{array}
\right) _1\left( 
\begin{array}{c}
1 \\ 
0
\end{array}
\right) _2,  \label{hu134}
\end{equation}

\begin{eqnarray}
\lbrack \Psi (1,0^{(\widehat{{\bf a}})};(m_1)^{(\widehat{{\bf c}}%
_1)},(m_2)^{(\widehat{{\bf c}}_2)})] &=&\frac 1{\sqrt{2}}\ \left( 
\begin{array}{c}
1 \\ 
0
\end{array}
\right) _1\left( 
\begin{array}{c}
0 \\ 
1
\end{array}
\right) _2  \nonumber \\
&&+\frac 1{\sqrt{2}}\left( 
\begin{array}{c}
0 \\ 
1
\end{array}
\right) _1\left( 
\begin{array}{c}
1 \\ 
0
\end{array}
\right) _2,  \label{hu135}
\end{eqnarray}
\begin{equation}
\lbrack \Psi (1,(-1)^{(\widehat{{\bf a}})};(m_1)^{(\widehat{{\bf c}}%
_1)},(m_2)^{(\widehat{{\bf c}}_2)})]=-\left( 
\begin{array}{c}
0 \\ 
1
\end{array}
\right) _1\left( 
\begin{array}{c}
0 \\ 
1
\end{array}
\right) _2  \label{hu136}
\end{equation}
and 
\begin{eqnarray}
\lbrack \Psi (0,0;(m_1)^{(\widehat{{\bf c}}_1)},(m_2)^{(\widehat{{\bf c}}%
_2)})] &=&\dfrac 1{\sqrt{2}}\left( 
\begin{array}{c}
1 \\ 
0
\end{array}
\right) _1\left( 
\begin{array}{c}
0 \\ 
1
\end{array}
\right) _2  \nonumber \\
&&-\dfrac 1{\sqrt{2}}\left( 
\begin{array}{c}
0 \\ 
1
\end{array}
\right) _1\left( 
\begin{array}{c}
1 \\ 
0
\end{array}
\right) _2.  \nonumber \\
&&  \label{hu137}
\end{eqnarray}

In this limit, the elements of $[r^{(1)}]_1$ are 
\begin{equation}
r_{11}^{(1)}=\cos ^2\theta _1/2r^{(1)}((+\frac 12)^{(\widehat{{\bf c}}%
_1)})-\sin ^2\theta _1/2r^{(1)}((-\frac 12)^{(\widehat{{\bf c}}_1)}),
\label{hu139a}
\end{equation}
\begin{equation}
r_{12}^{(1)}=\frac 12\sin \theta _1e^{-i\varphi _1}(r^{(1)}((+\frac 12)^{(%
\widehat{{\bf c}}_1)})-r^{(1)}((-\frac 12)^{(\widehat{{\bf c}}_1)})),
\label{hu139b}
\end{equation}
\begin{equation}
r_{21}^{(1)}=r_{12}^{(1)*}  \label{hu139c}
\end{equation}
and 
\begin{equation}
r_{22}^{(1)}=-r_{11}^{(1)}.  \label{hu139d}
\end{equation}
The elements of $[r^{(2)}]_2$ are 
\begin{equation}
r_{11}^{(2)}=\cos ^2\theta _2/2r^{(2)}((+\frac 12)^{(\widehat{{\bf c}}%
_2)})-\sin ^2\theta _2/2r^{(2)}((-\frac 12)^{(\widehat{{\bf c}}_2)})
\label{hu140a}
\end{equation}

\begin{equation}
r_{12}^{(2)}=\frac 12\sin \theta _2e^{-i\varphi _2}(r^{(2)}((+\frac 12)^{(%
\widehat{{\bf c}}_2)})-r^{(2)}((-\frac 12)^{(\widehat{{\bf c}}_2)}))
\label{hu140b}
\end{equation}

\begin{equation}
r_{21}^{(2)}=r_{12}^{(2)*}  \label{hu140c}
\end{equation}

\begin{equation}
r_{22}^{(2)}=-r_{11}^{(2)}  \label{hu140d}
\end{equation}

In the event that the quantities $r_1$ and $r_2$ are spin projections, we
may assign the values $+1$ if the projections are up with respect to the
respective unit vectors $\widehat{{\bf c}}_1$ and $\widehat{{\bf c}}_2$ and $%
-1$ if they are down with respect to these vectors. Thus, $r^{(1)}((\pm
\frac 12)^{(\widehat{{\bf c}}_1)})=\pm 1.$ In that case, the generalized
operator $[r^{(1)}]_1$ has the elements

\begin{equation}
r_{11}^{(1)}=\cos (\theta _d-\theta _1)-2\sin \theta _d\sin \theta _1\sin
^2((\varphi _d-\varphi _1)/2),  \label{hu141}
\end{equation}

\begin{equation}
r_{12}^{(1)}=-\sin \theta _d\cos \theta _1+\sin \theta _1\cos \theta _d\cos
(\varphi _d-\varphi _1)+i\sin \theta _1\sin (\varphi _d-\varphi _1),
\label{hu142}
\end{equation}

\begin{equation}
r_{21}^{(1)}=r_{12}^{(1)}  \label{hu143}
\end{equation}

and 
\begin{equation}
r_{22}^{(1)}=-r_{11}^{(1)}.  \label{hu144}
\end{equation}

Exactly the same expressions hold for the operator $[r^{(2)}]_2$, except
that the subscript $d$ is replaced by the subscript $f$, and where the
numeral $1$ does not give the row or column of a matrix element, it is
replaced by $2$.

In the limit $\widehat{{\bf d}}=\widehat{{\bf f}}=\widehat{{\bf k}}$, the
operators become

\begin{equation}
\lbrack r^{(1)}]_1=\left( 
\begin{array}{cc}
\cos \theta _1 & \sin \theta _1e^{-i\varphi _1} \\ 
\sin \theta _1e^{i\varphi _1} & -\cos \theta _1
\end{array}
\right) _1  \label{hu145}
\end{equation}
and 
\begin{equation}
\lbrack r^{(2)}]_2=\left( 
\begin{array}{cc}
\cos \theta _2 & \sin \theta _2e^{-i\varphi _2} \\ 
\sin \theta _2e^{i\varphi _2} & -\cos \theta _2
\end{array}
\right) _2,  \label{hu146}
\end{equation}
the well-known standard forms. Thus, we see that the standard results are
obtained easily and logically from this approach. These states are all
normalized to unity, as is easily proved. Also, they are mutually orthogonal.

\section{Discussion and Conclusion}

In this paper, we have derived the standard matrix treatment of spin
addition from probability amplitudes. This confirms the fact, first brought
out in Ref. [7], that spin theory can be based on probability amplitudes. It
also confirms the correctness of the Land\'e approach to quantum mechanics.

A very important observation arising from this paper is that the standard
results for spin matrix mechanics are only a special case of more
generalized ones. Despite that calculations can be successfully performed
with the standard quantities even in ignorance of this fact, our
understanding of spin theory is incomplete until we take this fact on board.
Although the results in this paper relate to spin addition, the general
observation that the standard theory of angular momentum addition is not
generalized enough is true. Therefore, more generalized results await the
application of the current approach to the addition of spins other than
those corresponding to spin-$1/2$ systems. By the same token, the addition
of spin and orbital angular momentum will lead to more generalized results.
One can extend this observation to the case of the addition of three or more
spins. The elucidation of angular momentum theory cannot be regarded as
complete until the task of obtaining the generalized results is finished.

From the generalized probability amplitudes derived in Ref. [1], we
presented two different matrix treatments for the triplets states, and one
matrix treatment for the singlet state. We found that we could express the
singlet state or the triplet states in terms of $4\times 4$ operators and
vectors with four rows each. In addition, we could express the triplet
states by means of $3\times 3$ operators and vectors with three rows. In
both cases, the total space was not decomposed into two spaces corresponding
to the constituent subsystems $1$ and $2$. But in the standard treatment,
the operator is the product of an operator in the space of subsystem $1$ and
of an operator in the space of subsystem $2$. The state consists of terms
which are products of vectors in the subspaces of systems $1$ and $2$. There
is thus this difference between the standard treatment and the new
treatments in Ref. [1]. This difference appears to be far from trivial. In
the present generalized standard treatment, we could only succeed in
deriving results by assuming that the operator of the arbitrary observable $%
R $ was factorizable into factors depending on the spaces of subsystem $1$
and of subsystem $2$. In the new treatments of Ref. [1], this was not
necessary. Thus, when we need to deal with ''entangled'' observables $R$, we
need to resort to the new treatments, or to forgo matrix mechanics and use
probability-amplitude mechanics.

Our work highlights the power of the Land\'e interpretation of quantum
mechanics. This approach continues to surprise, and it is all but certain
that it has new results to yield when applied to areas of quantum mechanics
other than spin theory.

\section{References}

1. Mweene H. V., ''New Treatment of Systems of Compounded Angular
Momentum'', quant-ph/9907082

2. Land\'e A., ''From Dualism To Unity in Quantum Physics'', Cambridge
University Press, 1960.

3. Land\'e A., ''New Foundations of Quantum Mechanics'', Cambridge
University Press, 1965.

4. Land\'e A., ''Foundations of Quantum Theory,'' Yale University Press,
1955.

5. Land\'e A., ''Quantum Mechanics in a New Key,'' Exposition Press, 1973.

6. Rose M. E., ''Elementary Theory of Angular Momentum'', John Wiley and
Sons, Inc. (New York), 1957

7. Mweene H. V., ''Derivation of Spin Vectors and Operators From First
Principles'', quant-ph/9905012

8. Mweene H. V., ''Generalized Spin-1/2 Operators and Their Eigenvectors'',
quant-ph/9906002

9. Mweene H. V., ''Vectors and Operators for Spin 1 Derived From First
Principles'', quant-ph/9906043

10. Mweene H. V., ''Alternative Forms of Generalized Vectors and Operators
for Spin 1/2'', quant-ph/9907031

11. Mweene H. V., ''Spin Description and Calculations in the Land\'e
Interpretation of Quantum Mechanics'', quant-ph/9907033

\end{document}